\documentclass[final,4p]{elsarticle}
\usepackage{amssymb}
\usepackage{graphicx}
\usepackage{multirow}
\usepackage{txfonts}
\usepackage{ragged2e}
\usepackage{etoolbox}
\usepackage{lipsum}
\begin{document}

\begin{frontmatter}

\title{Deuterium Enrichment of the Interstellar Medium}

\author[ICSP]{Ankan Das}
\ead{ankan@csp.res.in}
\author[ICSP]{Liton Majumdar}
\ead{liton@csp.res.in}
\author[SNBNCBS,ICSP]{Sandip K. Chakrabarti}
\ead{chakraba@bose.res.in}
\author[ICSP]{Dipen Sahu}
\ead{dipen@csp.res.in}
\address[ICSP]{Indian Centre For Space Physics, 43 Chalantika, Garia Station Road, Kolkata 700084, India}
\address[SNBNCBS]{S.N. Bose National Center for Basic Sciences, JD-Block, Salt Lake, Kolkata,700098, India}

\begin{abstract}
Despite low elemental abundance of atomic deuterium in interstellar medium (ISM), observational evidences suggest that several species 
in gas-phase and in ices could be heavily fractionated. We explore various aspects of deuterium enrichment by constructing a 
chemical evolution model in gas and grain phases. Depending on various physical parameters, gas and grains are allowed 
to interact with each other through exchange of their chemical species. It is known that HCO$^+$ and N$_2$H$^+$ 
are two abundant gas phase ions in ISM and their deuterium fractionation are generally used to predict degree of 
ionization in various regions of a molecular cloud. To have a more realistic estimation, we consider a density
profile of a collapsing cloud. We present radial distributions of important interstellar  molecules along with their 
deuterated isotopomers. We carry out quantum chemical simulation to study effects of isotopic substitution on spectral 
properties of these important interstellar species. We calculate vibrational (harmonic) frequency of the most important 
deuterated species (neutral \& ions). Rotational and distortional constants of these molecules are also computed to 
predict rotational transitions of these species. We compare vibrational (harmonic) and rotational transitions 
as computed by us with existing observational, experimental and theoretical results. We hope that our results 
would assist observers in their quest of several hitherto unobserved deuterated species.
\end{abstract}


\end{frontmatter}

\section{Introduction}
Study of deuterium enrichment received a major boost after discovery of singly or multiply deuterated H$_2$CO and CH$_3$OH in Interstellar Medium (ISM). 
Interestingly, fractionation ratios of these deuterated species often cross elemental D/H ratio of ISM ($\sim 1.5 \times 10^{-5}$, Linsky et al., 1995). 
Earlier work by Hasegawa, Herbst \& Leung (1992); Chakrabarti et al. (2006ab); Das et al. (2008a); 
Das, Acharyya \& Chakrabarti (2010); Cuppen \& Herbst (2007);  Das \& Chakrabarti (2011) 
suggest that grain chemistry plays a crucial role in deciding chemical composition of ISM, in general.
Role of grain chemistry towards deuterium enrichment has been highlighted also by various authors
(Caselli, 2002a; Cazaux et al., 2010; Das et al., 2013a).  
In gas phase, basic reactions are followed by dominant ion-molecular reaction pathways.
Around the cold, dense region of the cloud, CO and O are heavily depleted from the gas phase.
A strong correlation between the CO depletion and deuterium fractionation has been observed 
(Bacmann et al., 2003; Crapsi et al., 2005). Moreover, ionization of the ISM could be 
traced by observing some deuterated ions, such as, DCO$^+$ and N$_2$D$^+$ (Caselli, 2002a). 
So deuterium chemistry is extremely important for tracing dynamic properties of a cloud. 

Aikawa et al. (2005), Das et al. (2008b), and Das et al. (2013b) already discussed how 
dynamic parameters of a cloud affect its chemical composition.
Over the last two decades, several attempts were made to 
differentiate evolutionary stages of proto-stars. For example, Lada \& Wilking (1984) defined 
three classes of proto-stars (Class I, Class II and Class III)
which are progressively evolved. Andr\.e et al. (1993) defined another class of 
proto-stars, namely, Class 0, which is the youngest among protostars. 
When Class 0 proto-stars are in a little more evolved stage, the source is considered to be a 
Class 0/I borderline object. These stages are defined by
observing the variation of spectral energy distribution of protostars. A number of studies
are carried out till date to find out a connection between abundances of chemical species and 
evolutionary stages (e.g., Smith, 1998; Myers et al., 1998; Forebrich, 2005) of a proto-star. 
A number of past studies suggest that the deuterium fractionation of some ions like N$_2$H$^+$ and HCO$^+$ could 
be used to define different evolutionary stages (Crapsi, 2005; J$\phi$rgensen et al., 2004).
Recently, Majumdar et al. (2013) performed a quantum chemical calculation to obtain
spectral signatures (infrared and electronic absorption spectra) of precursors
of some bio-molecules such as adenine, alanine and glycine. It was found that 
spectral signatures of gas phase significantly differ from those in ice
phase. Das et al. (2013a) discuss different properties of HCOCN and one of its 
isotopologues, DCOCN. This type of studies could also be carried out for other deuterated species which could
serve as benchmarks for observation.
  
In this paper, we have presented a systematic approach to determine chemical evolution of some of the
most important deuterated species and tried to present a complete spectral catalog for detecting
these molecules around cold \& dense regions of a molecular cloud. 
The plan of this paper is as follow: In Section 2, models and 
computational details are presented. Implications of our results are discussed in
Section 3. Finally, in Section 4, we draw our conclusions. 

\section{Chemical Modeling}
\subsection{Gas phase Chemical Model}
We prepare a large gas-grain chemical network to iteratively study chemical processes
in a molecular cloud. Our gas phase chemical network consists of $6149$ reactions among $601$ species.
We mainly follow the UMIST-2006 database (Woodall et al., 2007) for construction of our
gas phase chemical network. Since in the present context, our motivation is to study
deuterium enrichment of the ISM, we use some of the important reactions from UMIST-2006 database 
and assume that these reactions would also be possible for the deuterated isotopomers as well.
Reaction rates are calculated by following the method used by Woodall et al. (2007).
To avoid long computational time as well as complexity in handling a
large chemical network, we identify some dominant pathways for deuterium enrichment and concentrate on them. Our selection is based on
earlier studies of Albertsson et al. (2013), Miller et al. (1989), Robert et al. (2000) and 
Rodgers \& Millar (1996). We assume that gas and grains are coupled through accretion and 
thermal/cosmic ray 
evaporation processes. Details of these processes are already presented in Das \& Chakrabarti (2011) and Das et al. (2013ab).

\subsection{Ice phase Chemical Model}

\subsubsection{Accretion}
Gas phase species are depleted by their accretion on interstellar ice. 
Following Hasegawa, Herbst \& Leung (1992), accretion rate of a gas phase species is given by:
$$
k_{acc} (i) = s_i \sigma v(i) n(i) \ s^{-1},
$$
where, $s_i$ is sticking coefficient, $n(i)$ is gas phase concentration and $v(i)$ is thermal velocity 
of $i^{th}$ species, $\sigma$ is geometrical dust-grain cross section ($\sigma=4\pi r^2$, r is the
radius of the grain$\sim 1000 \AA$).
In our simulation, $s_i=1$ is considered for all the neutrals except H$_2$ and He. 
It is not certain wheather atomic and molecular ions stick to the grain surfaces 
(Hasegawa, Herbst \& Leung, 1992; Watson, 1976), Here we considered $s_i=0$ for the ions.

\subsubsection{Binding energies}
Chemical enrichment of interstellar grain mantle solely depends on binding energies of surface species (Das \& Chakrabarti, 2013).
Mobility of lighter species such as H, D, N and O mainly dictates chemical composition of interstellar grain mantle. Composition of grain mantle 
which depends on mobility of H and O atoms are already discussed in Das et al. (2008a),
Das, Acharyya \& Chakrabarti (2010) and Das \& Chakrabarti (2011).
We assume that gas phase species are physisorbed onto dust grains ($\sim 0.1\mu$m) 
having a grain number density of $1.33 \times 10^{-12} n_H$, where $n_H$ is the concentration 
of H nuclei in all forms. Binding energies of deuterated species are assumed to be 
same as those of hydrogenated counterparts because binding energies of deuterated species are unknown.
Several theoretical and experimental attempts were made till date to find out
diffusion effects of atomic hydrogen. According to some past studies, as in 
Allen \& Robinson (1977), Tielens \& Allamandola (1987), Hasegawa \& Herbst (1993), 
Hasegawa, Herbst \& Leung (1992) and Chakrabarti et al. (2006ab), 
binding energy for diffusion (E$_b$) of 
H atom was considered to be $\sim 100$ K, whereas for desorption (E$_d$), 
it was considered to be $\sim 350$ K. Following Hasegawa \& Herbst (1993), for H$_2$ molecule, 
desorption energy was considered to be $\sim 450$ K. 
According to Hasegawa \& Herbst (1993) and references therein, adsorption energy of 
N$_2$ is $1210$ K. Caselli (2002a) and references therein, suggest 
that adsorption energy of N$_2$ could be $787$ K. In our simulation, we consider
$787$ K to be the adsorption energy for N$_2$. Desorption energies ($E_d$) for all other species
are taken from the past studies by Hasegawa \& Herbst (1993). Following Tielens and Allamandola (1987) 
and Hasegawa, Herbst \& Leung (1992), here also, we assume E$_b$=0.3E$_d$ for all other species except for 
H atom. To show importance of these binding energies towards chemical complexity of interstellar
grain mantle, we construct three sets of binding energies. First Set
consists of binding energies just mentioned above and we call it as set 1.
Unless otherwise stated, we always use set 1 energy values.
In set 2, we use results derived by Pirronello (1997, 1999) for
binding energies ($E_b$ and $E_d$) of H and H$_2$ with olivine grain surface.
So difference between set 1 and set 2 is that in set 2, we are using different binding
energies ($E_b$ and $E_d$ both) for H, H$_2$, D, HD and D$_2$ only.
Binding energies of all other species
are assumed to be similar to those belonging to the set 1. Similarly, in set 3, experimental findings of Pirronello
(1997, 1999) for binding energies of H and H$_2$ with the amorphous carbon grain are used.
For more clarity, in Table 1, we give these three sets of binding energies.
Since for all cases binding energies of all the species except 
H, D, H$_2$, HD, and D$_2$ are similar, we only tabulate the binding energies (E$_b$ and E$_d$) 
of these species only. For the sake of completeness in Table 1, we have shown the 
binding energies of some important surface species (O, OH, H$_2$O, CO, H$_2$CO, CH$_3$OH) as well.

\subsubsection{Reaction}
There are two reaction schemes, the Langmuir-Hinshelwood (LH) mechanism and
the Eley-Rideal (ER) mechanism normally considered for the surface reactions. 
In the LH scheme, the gas phase species accretes onto a grain and becomes equilibrated 
with the surface before it reacts with another atom or molecule, and in the ER reaction scheme, 
the incident gas phase species collides directly with an adsorbed species on the surface 
and reacts with that species.
In order to react, the adsorbed species require sufficient mobility. Surface
reaction rate $R_{i,j}$ between surface species $i$ and $j$ occurring due to classical diffusion
can be expressed as (Hasegawa et al., 1992),
$$
R_{i,j} = k_{i,j} (Rdiff_i + Rdiff_j )n_i n_j n_d ,
$$
where, $n_i$ and $n_j$ are the number of species $i$ and $j$ respectively, on an average
grain, $Rdiff_i$ and $Rdiff_j$ are the diffusion rate (defined as inverse of the diffusion
time), $k_{i,j}$ is the probability for the reaction to happen upon an encounter. The
parameter $k_{i,j}$ is in general unity for the exothermic reaction without activation
energy. For an exothermic reaction with activation energy $E_a$ and at least one light
reactant (H,H$_2$), $k_{i,j}$ can be approximated by the exponential portion of the quantum
mechanical probability for tunneling through a rectangular barrier of thickness $a$:
$$
k_{i,j} = exp[(-4\pi a/h)(2\mu E_a )^{1/2} ],
$$
where, $\mu$ is the reduced mass and $a$ is taken as $1\AA$.
For the light reactive species H and H$_2$, surface migration via tunneling is much
faster than that due to classical hopping. The time scale for tunneling is given by,
$$
t_{tun} = {\nu_0}^{-1} exp[(4 \pi a/h)(2mE_b )^{1/2} ] \ \ sec. 
$$
\begin{table*}
\centering{
\scriptsize
\caption{Various sets of binding energies}
\begin{tabular}{|c|c|c|c|c|c|c|}
\hline
Species&\multicolumn{2}{|c|}{set 1}&\multicolumn{2}{|c|}{set 2}&\multicolumn{2}{|c|}{set 3}\\
\cline{2-7}
&$E_b$&$E_d$&$E_b$&$E_d$&$E_b$&$E_d$\\
\hline
\hline
H&100&350&287&373&511&657\\
H$_2$&135&450&95&315&163&542\\
D&100&350&287&373&511&657\\
HD&135&450&95&315&163&542\\
D$_2$&135&450&95&315&163&542\\
O&240&800&240&800&240&800\\
OH&378&1260&378&1260&378&1260\\
H$_2$O&558&1860&558&1860&558&1860\\
CO&363&1210&363&1210&363&1210\\
H$_2$CO&528&1760&528&1760&528&1760\\
CH$_3$OH&618&2060&618&2060&618&2060\\
\hline
\end{tabular}}
\end{table*}

For the grain surface reaction network, we primarily follow Hasegawa, Herbst \& Leung (1992),
Cuppen \& Herbst (2007), Das, Acharyya \& Chakrabarti (2010) and Das \& Chakrabarti (2011). 
For deuterium fractionation reaction on the grain surface, we primarily follow Caselli (2002a) 
and Cazaux et al. (2010).

\subsubsection{Thermal evaporation}
In our model, abundances of surface species could be decreased by thermal evaporation,
cosmic ray induced evaporation and non-thermal desorption processes. 
Rate of thermal evaporation of the surface species `i' could be calculated by the following relation,
\begin{equation}
k_{evap}(i)=\nu_0 exp(-E_d/kT_g) \ \ sec^{-1},
\end{equation}
where, $\nu_0$ is characteristic vibrational frequency ($\nu_0=\sqrt{2 n_s E_d/\pi^2 m}$) and 
$T_g$ is temperature of grain. For all cases, we have considered 
that total number of sites ($N_S$) on a grain is $10^6$ having surface density 
of sites (n$_s$) of $2 \times 10^{14}$ cm$^{-2}$.

\begin{table*}
\centering{
\caption{Initial abundances}
\begin{tabular}{|c|c|}
\hline
Species&Abundance\\
\hline
\hline
H$_2$ &    $5.00 \times 10^{-01}/n_H$\\
He    &    $1.00 \times 10^{-01}/n_H$\\
N     &    $2.14 \times 10^{-05}/n_H$\\
O     &    $1.76 \times 10^{-04}/n_H$\\
H$_3$$^+$&    $1.00 \times 10^{-11}/n_H$\\
C$^+$ &    $7.30 \times 10^{-05}/n_H$\\
S$^+$ &    $8.00 \times 10^{-08}/n_H$\\
Si$^+$&    $8.00 \times 10^{-09}/n_H$\\
Fe$^+$&    $3.00 \times 10^{-09}/n_H$\\
Na$^+$&    $2.00 \times 10^{-09}/n_H$\\
Mg$^+$&    $7.00 \times 10^{-09}/n_H$\\
P$^+$ &    $3.00 \times 10^{-09}/n_H$\\
Cl$^+$&    $4.00 \times 10^{-09}/n_H$\\
e$^-$ &    $7.31 \times 10^{-05}/n_H$\\
HD&  $1.60 \times 10^{-05}/n_H$\\
H&  $1.00 \times 10^{00} \ cm^{-3}$\\
D&  $1.00 \times 10^{-01} \ cm^{-3}$\\

\hline
\end{tabular}}
\end{table*}

\subsubsection{Cosmic ray induced evaporation}
Cosmic ray induced evaporation (hereafter CRD) is a very efficient means to transfer surface 
molecules into gas phase during late stage of chemical evolution. Cosmic ray induced 
evaporation rates are calculated by using the expression developed by Hasegawa \& Herbst (1993).
Following Leger et al. (1985), they assumed that relativistic Fe nuclei with
energies $20-70$MeV could deposit $0.4$MeV energy on an average dust particle of
radius $0.1\mu$m. Grains could be cooled down due to thermal evaporation and radiation processes.
For easy inclusion of cosmic ray induced photo-evaporation into their model, they developed the 
following relation:
\begin{equation}
k_{crd}\sim f(70,K) k_{evap}(i,70 \ K),
\end{equation}
where, $k_{evap}(i, 70 \ K)$ is the thermal evaporation rate of surface species `i' at temperature $70$ K,
$f(70 \ K)$ is fraction of time spent by grains at around
$70$ K. Following Leger et al. (1985), they defined $f(70 \ K)=3.16 \times 10^{-19}$.

\subsubsection{Non thermal desorption}
Because energy is released during some of the reactions, adsorbed species could desorb 
just after their formation. Garrod et al. (2007) estimated desorption rate via exothermic surface 
reactions by considering Rice-Ramsperger-Kessel (RRK) theory. 
They parameterized non-thermal desorption by assuming 
some approximation. They assumed that a fraction `f' of the product species in qualifying 
reactions could desorb immediately and the rest (1-f) fraction remains as a surface bound product.
Here, we apply this mechanism to all surface reactions which result in a single product. 
Fraction `f' is calculated by;
\begin{equation}
f=\frac{aP}{(1+aP)},
\end{equation}
where, $a$ is ratio between surface molecule bond frequency to frequency at which energy is lost to
grain surface. Garrod et al. (2007), adopted similar $a$ values for all the species. 
They varied $a$ from $0.01$ to $0.1$. 
A value of $a = 0.1$ was labeled as `high' by Pilling (2006). Kroes \& Andersson (2006) carried out 
molecular dynamics simulations of the irradiation of water ice with UV photons. From their data, 
Garrod et al. (2007) estimated that $\sim 0.9\%$ of recombinations result in desorption. 
Using the  value of $E_D (H2O) = 5700$ K and $E_{reac} = 5.91 \times 10^4$ K, they had  $a = 0.012$. 
Recently, Dulieu et al. (2013) experimentally found that 90 \%  of H$_2$O formed on surfaces 
with OH $+$ H is directly released into the gas phase (Dulieu et al. 2013). But this is yet to 
be verified for the other species. Here we are considering a huge gas-grain chemical network. 
In order to test the effects of the non-thermal desorption mechanism and constrain the value of $a$, 
we investigate models with various values of $a: 0, 0.01, 0.05, 0.1$. To be on the safer side, 
we choose an intermediate value of $a$ ($a=0.05$) for all surface reactions.

\begin{table}
\addtolength{\tabcolsep}{-2pt}
\caption{\bf Fractionation ratios of the Ice phase species for various sets of energies}
{\scriptsize
\begin{tabular}{|c|c|c|c|c|c|}
\hline
{\bf Species}&{\bf Isotopomers}&{\bf Fractionation ratio}&{\bf Fractionation ratio}&{\bf Fractionation Ratio}&{\bf Fractionation ratio}\\
&&{\bf (column density in cm$^{-2}$)}&{\bf (column density in cm$^{-2}$)}&{\bf (column density in cm$^{-2}$)}&{\bf (column density in cm$^{-2}$)}\\
&&{\bf by using set 1}&{\bf by using set 2}&{\bf by using set 3}&{\bf by using}\\
&&&&&{\bf experimental activation barrier}\\
&&&&&{\bf along with set 1}\\
\hline\hline
&HDO&$9.93 \times 10^{-02}$($1.62 \times 10^{17}$)&$1.97\times 10^{-01}$$(2.47 \times 10^{17})$&$2.11\times 10^{-02}$$(4.60\times 10^{16})$&$6.99\times 10^{-02} (1.40 \times 10^{17} )$\\
H$_2$O&D$_2$O&$1.39 \times 10^{-03}$($2.26 \times 10^{15}$)&$1.10\times 10^{-04}$$(1.38 \times 10^{14})$&$9.42\times 10^{-05}$$(2.05\times 10^{14})$&$5.72\times 10^{-04}(1.15 \times 10^{15})$\\
\hline&HDCO&$5.37 \times 10^{00}$$(2.89 \times 10^{10})$&$2.08\times 10^{-01}$$(1.35 \times 10^{15})$&$5.35\times 10^{-02}$$(2.61\times 10^{16})$&$4.84 \times 10^{02}(9.53 \times 10^{07})$\\
H$_2$CO&D$_2$CO&$8.24\times 10^{-02}$($4.43 \times 10^{08}$)&$7.46\times 10^{-04}$$(4.84 \times 10^{12})$&$5.40\times 10^{-04}$$(2.63\times 10^{14})$&$1.10\times 10^{02}(2.16 \times 10^{05})$\\
\hline
&CH$_3$OD&$3.06\times 10^{-02}$($1.11 \times 10^{16}$)&$4.19\times 10^{-03}$$(1.24 \times 10^{15})$&$9.39\times 10^{-03}$$(5.22\times 10^{11})$&$1.01\times 10^{-02}(6.06 \times 10^{15})$\\
&CH$_2$DOH&$8.24\times 10^{-02}$$(2.98\times 10^{16})$&$8.84\times 10^{-02}$$(2.62 \times 10^{16})$&$6.12\times 10^{-02}$$(3.40\times 10^{12})$&$1.14 \times 10^{-02}(6.85 \times 10^{15})$\\
&CHD$_2$OH&$1.09\times 10^{-03}$$(3.94 \times 10^{14})$&$2.55\times 10^{-04}$$(7.55 \times 10^{13})$&$7.11\times 10^{-04}$$(3.95\times 10^{10}$&$4.59 \times 10^{-02}(2.75 \times 10^{16})$\\
CH$_3$OH&CD$_3$OH&$1.06\times 10^{-07}$$(3.82 \times 10^{10})$&$4.59\times 10^{-09}$$(1.36 \times 10^{09})$&$3.97\times 10^{-07}$$(2.20\times 10^{07})$&$2.43 \times 10^{-04}(1.45 \times 10^{14})$\\
&CD$_3$OD&$3.11\times 10^{-09}$$(1.13 \times 10^{09})$&$2.10\times 10^{-11}$$(6.24 \times 10^{06})$&$2.88\times 10^{-10}$$(1.60\times 10^{04})$&$2.98 \times 10^{-06}(1.78 \times 10^{12})$\\
&CH$_2$DOD&$2.14\times 10^{-03}$$(7.73 \times 10^{14})$&$3.69\times 10^{-04}$$(1.10 \times 10^{14})$&$9.52\times 10^{-02}$$(1.10\times 10^{16})$&$2.51 \times 10^{-04}(1.50 \times 10^{14})$\\
&CHD$_2$OD&$2.71\times 10^{-05}$$(9.80 \times 10^{12})$&$1.16\times 10^{-06}$$(3.44 \times 10^{11})$&$6.60\times 10^{-06}$$(3.67\times 10^{08})$&$4.97 \times 10^{-04}(2.98 \times 10^{14})$\\
\hline
\end{tabular}}
\end{table}

\subsection{Results of chemical modeling}

\begin{figure}
\vskip 1cm
\centering{
\vbox{
  \includegraphics[height=8cm,width=12cm]{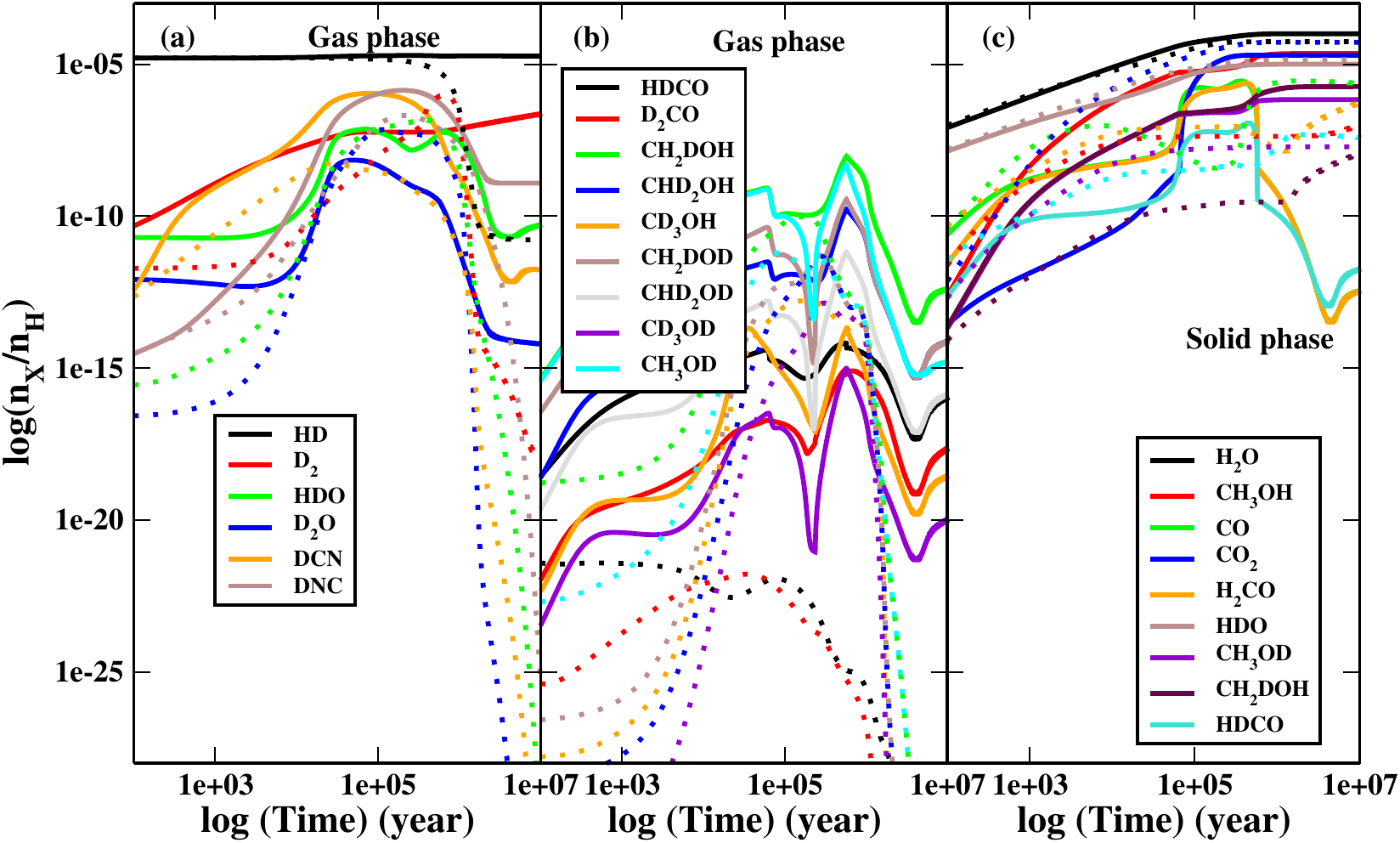}}}
\caption{\small Chemical evolution of some important gas(a,b) and ice(c) phase neutral deuterated species in ISM.
Solid lines are for cases where CRD and non-thermal desorption effects are considered and
dotted lines are for cases where these two effects are not considered.}
\end{figure}

Hydrogen molecules are mainly formed in diffuse regions evolving into molecular clouds through surface chemistry.
In this study, we aim at studying chemical evolution of a molecular cloud into dense 
cores. For this purpose, we adopt initial conditions from molecular clouds, where almost all 
hydrogens are in the form of molecular hydrogen, and temperature of gas and dust are 
assumed to be $\sim 10$ K. Ionization in these regions is mainly governed by Cosmic rays.
We adopt an initial condition by following Leung, Herbst \& Huebner (1984) (Table 2). They considered two sets of 
initial conditions (i) when metalicity is high and (ii) when metalicity is low. 
For set (ii) S, Si, Na, Mg and Fe are assumed to be depleted by two orders of magnitude as compared to set (i).
Leung, Herbst \& Huebner (1984) found that results of set (i) 
agree better with observed molecular abundances in cool, dense clouds. Here also we 
are considering a similar initially low metalicity condition. It is assumed that initially all 
deuteriums could be locked in the form of HD (just as we assumed that all H are in the form of
hydrogen molecules initially). Initial abundance of HD molecule is assumed to 
be $1.6 \times 10^{-5}$ with respect to total hydrogen nuclei. Initial abundance of atomic form of hydrogen and deuterium are considered to be 
$1$  $cm^{-3}$ and $0.1 \ cm^{-3}$ respectively. 
Unless otherwise stated, we always use initial abundance given in Table 2.

\subsubsection{Neutral deuterated species}
In Fig. 1ab, chemical evolution of some important gas phase neutral deuterated species 
with respect to total hydrogen ($n = n_H + 2 \times {n_H}_2$) are shown. Here, we assume 
that the cloud is at $T=10$ K having constant number density ($n = 10^4$  cm$^{-3}$), visual extinction 
$A_v=10$. Solid lines of Fig. 1ab, represent chemical evolution of gas phase species, 
where we have considered CRD and non-thermal desorption effects. Dotted lines are for 
the cases where we have not considered CRD and non-thermal desorption effects. 
As expected, CRD and non-thermal desorption serve as efficient means to maintain a reasonable 
gas phase abundance during later stages of chemical evolution process.   
Due to depletion  of gas phase deuterated species, curves show a decreasing trend after 
achieving a peak value. Solid lines show that gas phase species are maintained in a steady state 
equilibrium because of the effects of CRD and non-thermal desorption. 
Figure 1a clearly shows that HD molecule is the most abundant deuterated
gas phase species throughout the evolution. D$_2$ is also found to be significantly abundant. 
Initially, all deuteriums were locked in the form of HD molecule. By virtue of
low binding energies with grain surface, HD and D$_2$  maintain high abundances 
during late stages of chemical evolution process.
Two isotopomers of H$_2$O (HDO and D$_2$O) and HCN (DCN and DNC) are found
to be significantly abundant in gas phase. 
Chemical evolution of several isotopomers of CH$_3$OH and H$_2$CO are shown in Fig. 1b. 
Singly deuterated isotopomers of methanol (CH$_3$OD, CH$_2$DOH) are found to be reasonably abundant.
Along with doubly and triply deuterated isotopomers, tetra deuterated isotopomer is found to be 
produced in gas phase as well. Abundance of HDCO and D$_2$CO are also significant 
while we are considering CRD and non-thermal desorption effects.
These molecules are most likely formed on dust surfaces during cold, dense 
pre-collapse period and are evaporated to the gas phase (Loinard et al., 2000).
From Fig. 1ab, it is clear that when CRD and non-thermal desorption effects are taken into 
account, abundances of gas phase species reach  a steady state equilibrium beyond $10^6$ year. 
Species attain their peak values $\sim 1 \times 10^5$ year.
Depending on physical condition (density and temperature), accretion rate of the
gas phase species changes. As per Caselli (2002a), for a normal dust to gas ratio
and assuming typical dust grains with radius of $0.1\mu$m, the {\it average} time scale for a gaseous
species to be deposited onto a grain is
$$
t_D \sim 10^9 \sqrt(A_X) /[S n(H_2)] \ \mathrm{year}, 
$$
where, $A_X$ is molecular weight, $S$ is sticking coefficient and $n(H_2)$ is number
density of molecular hydrogen. Sticking coefficient of all neutral species are assumed to be unity.
Results presented in Fig. 1ab are for the number density $(n_H)= 10^4$ cm$^{-3}$ 
(i.e., for $n_{H_2}=5 \times 10^3$ cm$^{-3}$).
From the above equation the time scale for depleting species from the gas is 
$\sim 1 \times 10^5$ years. 
As depletion is inversely proportional to the density, for high density cloud, 
depletion time would be much shorter.

In Fig. 1c, chemical evolution of the most abundant surface species (H$_2$O, CO, CO$_2$, H$_2$CO, CH$_3$OH)
along with their most abundant deuterated analogues (HDO, HDCO, CH$_3$OD, CH$_2$DOH) 
are shown. As gas phase species deplete in around $2 \times 10^5$ years, we
are obtaining a steady state for the abundances of surface species beyond that time. 
Singly deuterated water, formaldehyde and methanol are found to be heavily abundant. 
From Fig. 1b, significant difference between gas phase abundances of these species for two cases 
(with CRD and non-thermal desorption process and without these effects) are distinctly visible.
Gas-phase abundances of species like methanol and formaldehyde cannot 
be explained by gas-phase chemistry alone. In fact, the sole potential synthesis of these species 
seems to be production on surfaces of interstellar grains followed by desorption into the gas. 
Yet, thermal evaporation process (binding energies of these species with grain surfaces are much higher
$\sim 2000$ K) is inefficient for explaining abundances of 
molecules such as methanol and formaldehyde at around $10$ K. It
is necessary then to contemplate non-thermal mechanisms as natural processes.
Formaldehyde and methanol are mainly formed by hydrogenation reaction of CO on interstellar dust
grains, and are released in the gas phase in hot core regions.
Noble et al. (2012) recently performed experiments to study various
desorption processes. Another interesting point to be noted from Fig. 1c is that 
surface abundances of CH$_3$OH and its deuterated isotopomers
increase when we consider non-thermal desorption process (solid lines).
It was expected that inclusion of non-thermal desorption parameter will increase
gas phase abundances of these species by virtue of decrease of surface abundances of these
species. But as in Garrod et al. (2007), here also we notice a strong enhancement of 
surface abundances of methanol along with its deuterated isotopomers. 
Garrod et al. (2007) mentioned that when non-thermal desorption effect were
not considered,  most of the surface methanol channeled into CH$_4$ at late times.
But when non-thermal desorption process was considered, carbon hydrides are allowed to return to gas phase,
where they could convert into CO or its hydrides and maintain 
a modest level of CO, formaldehyde and methanol in both gas phase and on grain surfaces. 
In order to study effects of non-thermal desorption on these species,
we assumed various `$a$' values and have separately presented 
results in Fig. 2. Three curves are shown for formaldehyde and methanol with
$a=0$, $a=0.01$ and $a=0.1$. As expected, for higher `$a$' values, gas phase abundances of 
formaldehyde and methanol are enhanced. During late stages of 
chemical evolution, gas phase abundance of these species are significant due to 
non-thermal desorption process. So it is essential to consider non-thermal desorption process 
into any gas-grain chemical model especially for low temperature regime. 
From now on, we consider the intermediate value of `$a$', namely, $a=0.05$.

Chemical composition resulting from our model solely depends on initial abundances
considered. In order to test this feature, we consider a similar physical 
condition for carbon rich environment. Unless otherwise stated, we  always consider
initial abundances of C$^+$ to be $7.3 \times 10^{-5}$. 
To mimic a carbon rich environment, we assume high carbon 
abundance initially ($1.4 \times 10^{-4}$ relative to hydrogen nuclei) in ionized (C$^+$) and 
in the neutral (C) form. For the purpose of illustration, chemical evolution of HCO$^+$ is  
shown in Fig. 3. Abundance of HCO$^+$ vary significantly with the choice of 
initial form (ionized or neutral) and abundance of carbon.
For higher initial C$^+$ abundance, we have much higher HCO$^+$ as expected and for 
higher neutral carbon abundance, we have much lower HCO$^+$ abundance as expected.

\begin{figure}
\vskip 1cm
\centering{
\vbox{
  \includegraphics[height=8cm,width=10cm]{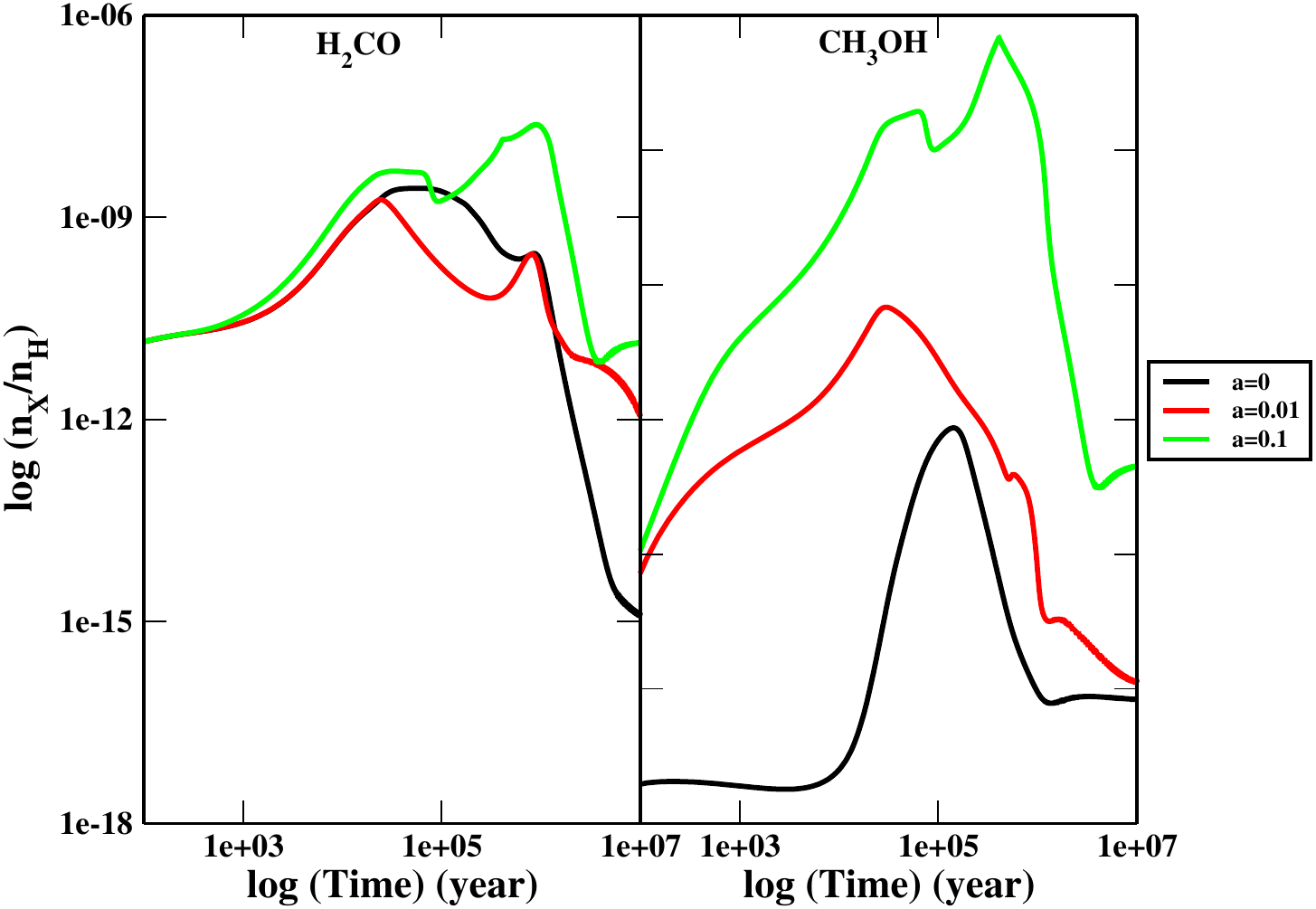}}}
\caption{\small Chemical evolution of gas phase H$_2$CO and CH$_3$OH are shown for
various values of `$a$'.}
\end{figure}

\begin{figure}
\vskip 1cm
\centering{
\vbox{
  \includegraphics[height=7cm,width=7cm]{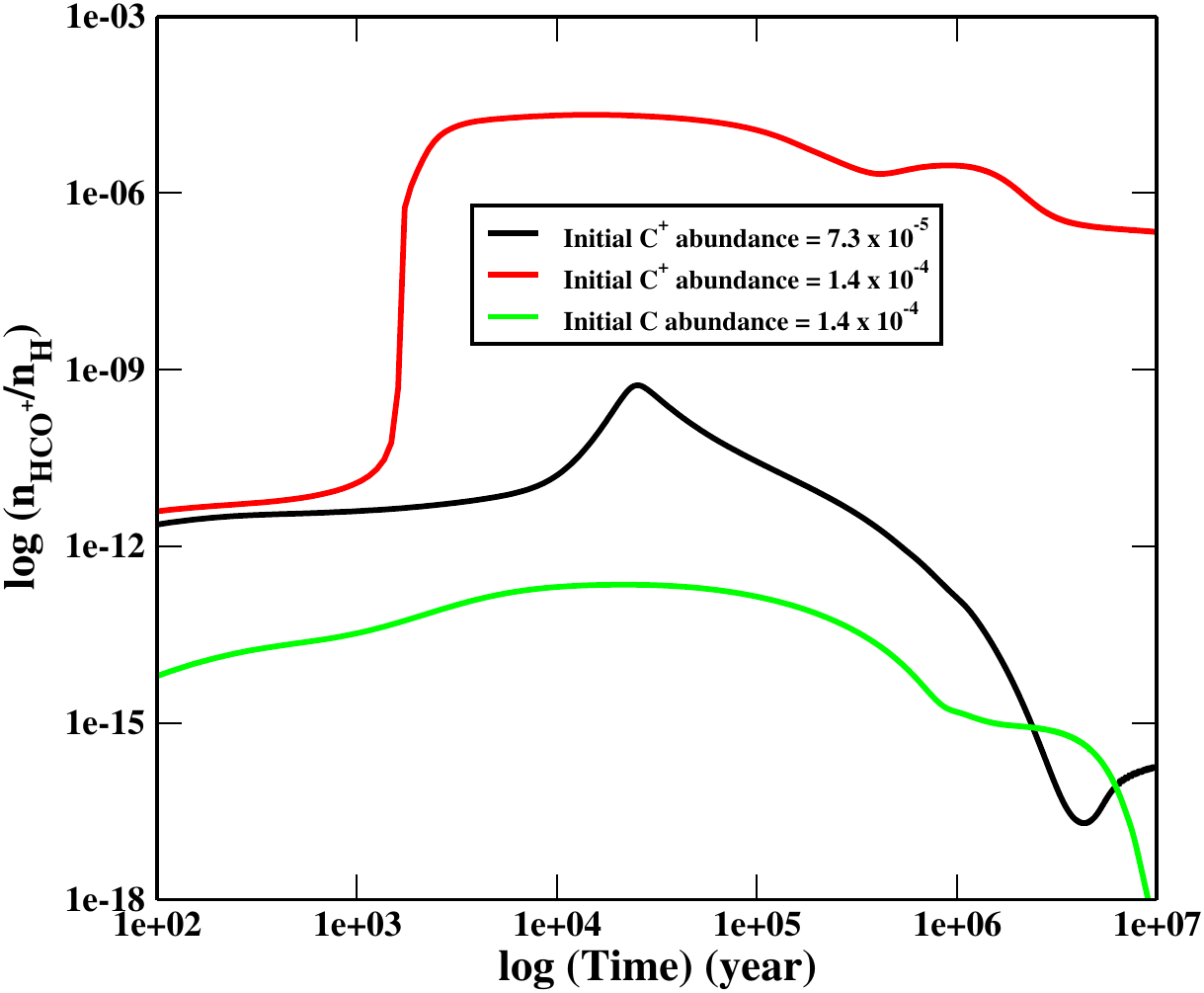}}}
\caption{\small Chemical evolution of gas phase HCO$^+$ for various initial carbon abundances.}
\end{figure}

\begin{figure}
\vskip 1cm
\centering{
\vbox{
\includegraphics[height=7cm,width=7cm]{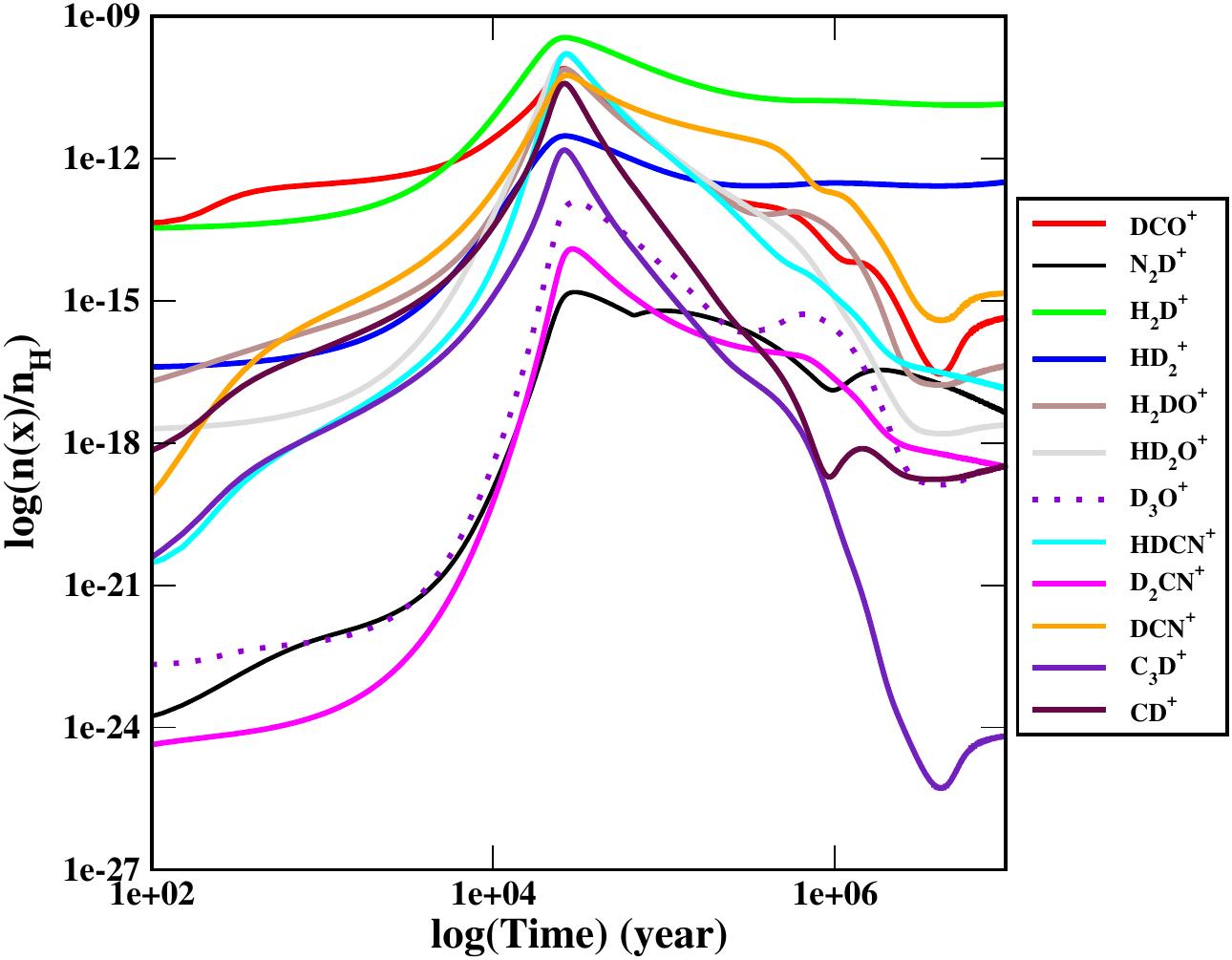}}}
\caption{\small Important deuterated molecular ions in gas phase.}
\end{figure}

\subsubsection{Deuterated ions}

In Fig. 4, we show chemical evolution of the most important deuterated molecular ions;
DCO$^+$, N$_2$D$^+$, H$_2$D$^+$, HD$_2^+$, H$_2$DO$^+$, HD$_2$O$^+$, D$_3$O$^+$, HDCN$^+$,
D$_2$CN$^+$, DCN$^+$, C$_3$D$^+$ and CD$^+$. Figure 4 clearly shows that beyond $2 \times 10^5$ years, 
abundances of these gas phase deuterated ions decrease gradually. 
Due to depletion of neutrals in around $2 \times 10^5$ years, production of related ions 
are also heavily hindered. Since recombination rate coefficients (reaction with electrons) of these ions
are much higher ($ \sim 10^{-6}$ to $10^{-7}$ cm$^3$s$^{-1}$), destruction rate of ions is much 
faster than the production rate. As a result, ions are disappearing at the same time 
scale as the neutrals. Among these ions, DCO$^+$ and N$_2$D$^+$ are widely used to correlate 
degree of ionization of ISM. Figure 4 shows that DCO+ molecule is very abundant in the gas phase.
The ratio DCO$^+$/HCO$^+$ (hereafter $R_1$) is widely used as a measure of limit on electron
abundances (x$_e$) in ISM. According to Roberts \& Millar (2000),
at low temperatures, DCO$^+$ is primarily formed via $H_2D^+ + CO \rightarrow DCO^+ + H_2.$
But there are other efficient routes for the formation of this species.
Here, we have followed pathways mentioned in Albertson et al. (2013) and
Robert \& Millar (2000) for production and destruction of DCO$^+$.
According to Albertsson et al. (2013), major formation routes of DCO$^+$
are the followings;

$$
\mathrm{H_2D^+ + CO \rightarrow DCO^+ + H_2}
$$
$$
\mathrm{HCO^+ + D\rightarrow DCO^+ + H}
$$
$$
\mathrm{{D_3}^+ + CO \rightarrow DCO^+ + D_2}
$$
$$
\mathrm{DOC^+ + H_2 \rightarrow DCO^+ + H_2}
$$
$$
\mathrm{CH_2D^+ + O \rightarrow DCO^+ + H_2}
$$
and major destruction routes are the followings,

$$
\mathrm{DCO^+ + e^- \rightarrow CO + D}
$$
$$
\mathrm{DCO^+ + SO\rightarrow DSO^+ + CO}
$$
$$
\mathrm{DCO^+ + H \rightarrow HCO^+ + D}
$$
$$
\mathrm{DCO^+ + C \rightarrow CD^+ + CO}
$$
$$
\mathrm{DCO^+ + HCN \rightarrow HDCN^+ + CO}
$$
$$
\mathrm{DCO^+ + HNC \rightarrow HDCN^+ + CO}
$$

\begin{figure}
\vskip 1cm
\centering{
\vbox{
\includegraphics[height=8cm,width=8cm]{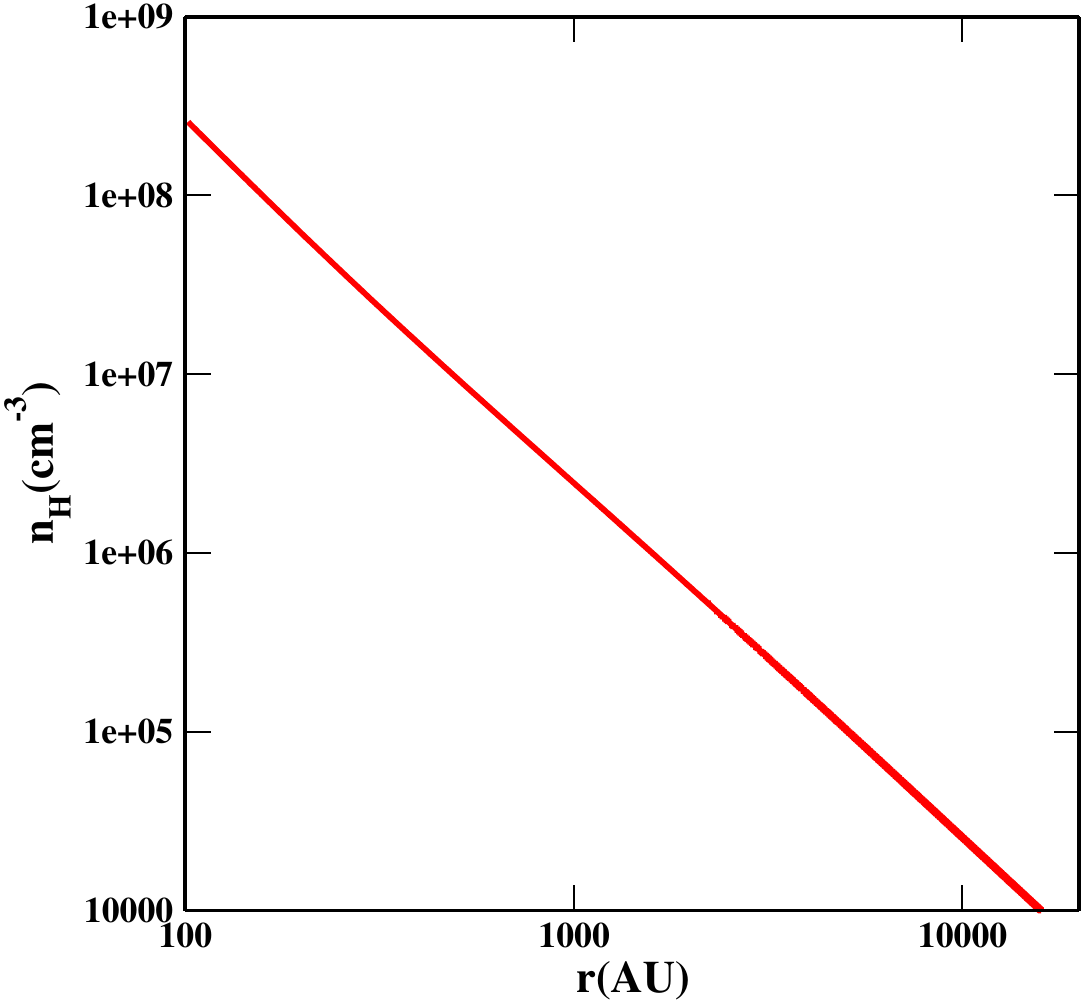}}}
\caption{\small Density profile of a collapsing cloud.}
\vskip 1cm
\end{figure}
\subsubsection{Modeling for a Molecular cloud}
To have  results under a more realistic situation, and to have radial distribution
of various interstellar species, we now consider that density profile follows
$\rho \sim r^{-2}$ distribution as described by Shu (1977).
Here we consider that inner boundary of protostar is at 100AU and outer boundary was chosen 
in such a way that density would become $10^4 cm^{-3}$.
In Fig. 5, density profile of the collapsing cloud is shown.
\begin{figure}
\centering{
\vbox{
\includegraphics[height=7cm,width=7cm]{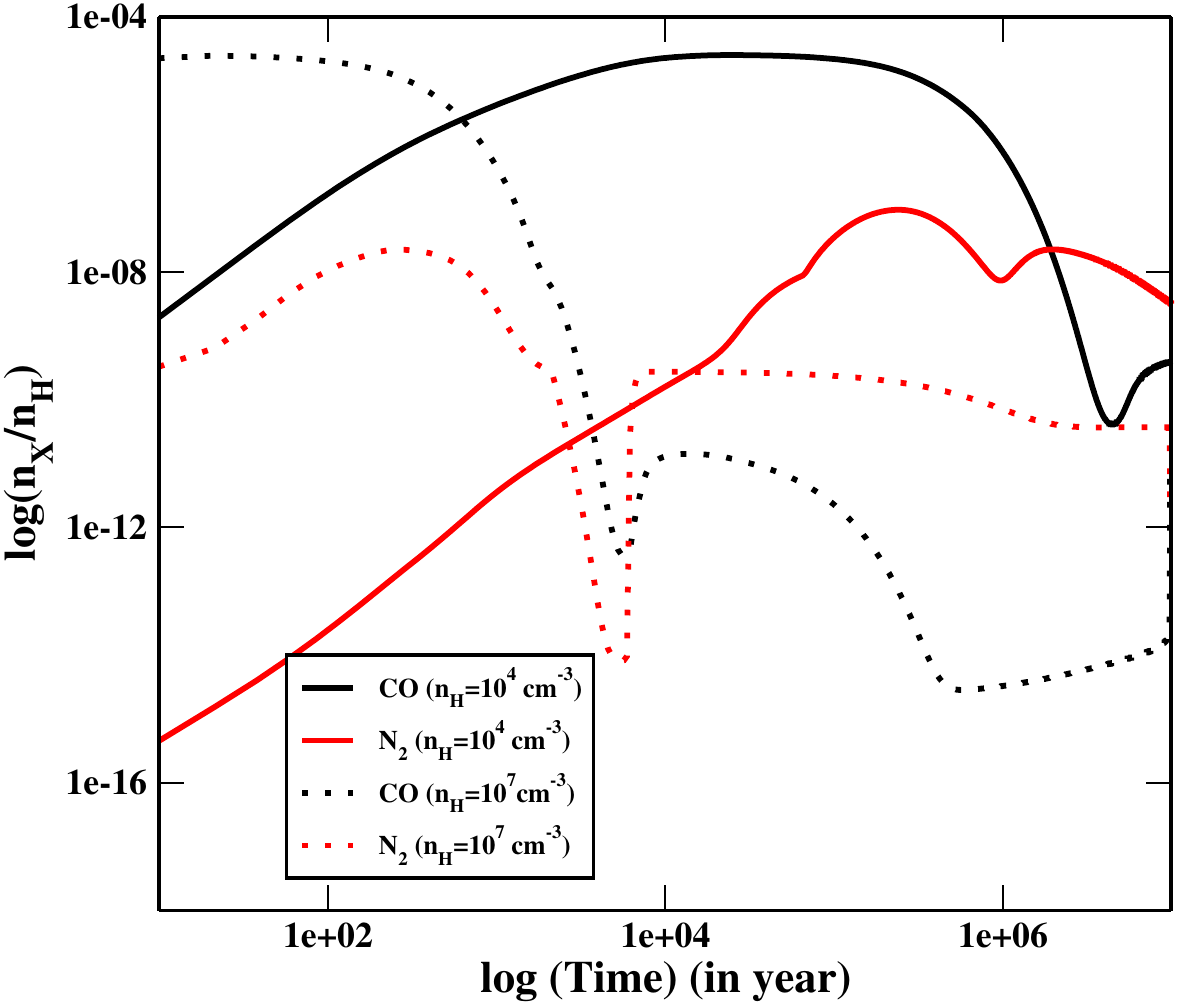}
\caption{\small Chemical evolution of gas phase CO and N$_2$ for n$_H=10^4$ cm$^{-3}$ and $n_H=10^6$ cm$^{-3}$.}}}
\end{figure}
\begin{figure}
\centering{
\vbox{
\includegraphics[height=9cm,width=9cm]{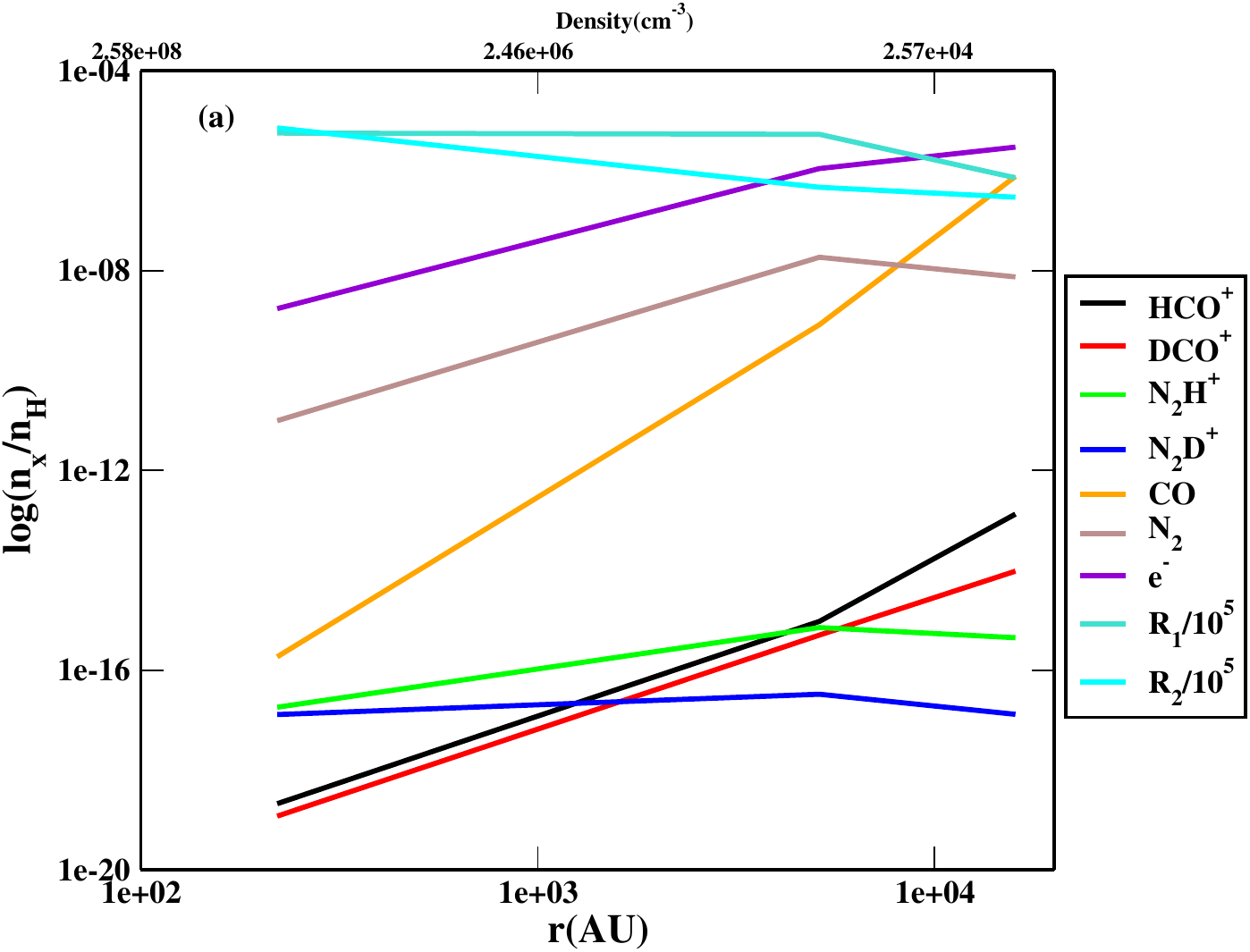}
\caption{\small Radial distribution of some important gas phase ions, R$_1$,  R$_2$ along with CO, N$_2$ and e$^-$. }}}
\end{figure}
In our simulations, we evolve our chemical code for various values of number densities
of total hydrogen nuclei ($n_H$) up to 
life time of a generic molecular cloud ($\sim 10^7$ years). 
For the sake of simplicity, we assume that from the beginning, ($t=0$) this cloud 
maintains this density profile. Initial conditions are assumed to be similar to the 
initial conditions of a molecular cloud. We picked up different density at various regions of the 
cloud (form Fig. 5) and studied chemical evolution of these regions.
For all cases, it was assumed that $T=10$ K and $A_V=10$.
 
In order to test depletion time scale of CO and N$_2$, in Fig. 6, we show
chemical evolution of CO and N$_2$ at two different density regions of the cloud. Solid lines are for 
${n_H}=10^4$ cm$^{-3}$ and dotted lines are for $n_H=10^7$ cm$^{-3}$. 
For both cases, CO is depleted much earlier than N$_2$.
In case of $n_H=10^4$ cm$^{-3}$, around $\sim 2 \times 10^6$ year CO molecules 
are heavily depleted whereas for $n_H=10^7$ cm$^{-3}$, 
this depletion time scale is shifted by about $\sim 8 \times 10^3$ year. 
In case of N$_2$, depletion features were obtained but it maintains higher abundances
during the later stages. This is mainly because of
low binding energy of N$_2$ ($E_d=787$ K) in comparison to that of CO ($E_d=1210$ K) with grain surfaces.
Deuterated species are normally used to
trace slightly warmer region of the core. R$_1$ is normally used to
co-relate ionization degree ($n_e/n_H$) of ISM. But species like CO could be depleted
at their dense interior (clearly visible from Fig. 6) and could severely affect R$_1$.
In order to justify appropriate deuterium fractionation of an entire core,
one has to observe molecules that are less affected by depletion, such as N$_2$H$^+$ and
NH$_3$. Another benefit of observing N-bearing molecules is that
their emission lines are splited into hyperfine components due to
non-zero nuclear spin of nitrogen, enabling optical depths of intervening medium to be
measured. So it is customary to study the ratio N$_2$D$^+$/N$_2$H$^+$ (hereafter R$_2$)
to have a better approximation around that region.

Figure 7 shows abundances of HCO$^+$, DCO$^+$,
${\mathrm N_2H^+}$ and ${\mathrm N_2D^+}$ at various depths. Abundances are taken from our
simulation results at $t=10^6$ year. 
From Fig. 7, it is clear that as density increases inside, abundances of all gas phase ions
decreases. In Fig. 6, we have shown that N$_2$ molecules deplete more slowly 
due to lower binding energies. A similar thing is also seen in Fig. 7.
Figure 7 shows that deep inside the cloud CO depletes heavily. N$_2$ is also
depleted but its rate of depletion is much slower in comparison to CO. As a result,
CO related species (HCO$^+$ and DCO$^+$) are depleting heavily, whereas
N$_2$ related species (N$_2$D$^+$ and N$_2$H$^+$) depletes very slowly around the dense interior.
Radial distributions of R$_1$ and R$_2$ along with the abundance of e$^-$
are also shown in Fig. 7. From Fig. 7, it is noted that R$_1$ remains roughly constant deep 
inside, while R$_2$ steadily increases with depth. This implies that as we 
go inside the cloud and density increases, deuterium fractionation of N$_2$H$^+$ becomes
favourable. However, in case of HCO$^+$, deuterium fractionation roughly remains constant. 

In Fig. 8, we plot 
electron abundance (a) with respect to R$_1$ and (b) with respect to R$_2$.  
It is already mentioned that due to heavy depletion of CO and its related species 
around the dense interior, N$_2$ related species should be used 
to predict the degree of ionization ($n_e/n_H$). We divide the entire zone of 
molecular clouds in two parts: first one is the outer shell (extended from 6000AU to 20000AU) and
second one is the dense interior (extended from 100AU to 6000 AU).
$R_1$ could be used to trace degree of ionization of the outer shell. From our model,
we have electron abundance in the range of $10^{-6}-3 \times 10^{-6}$. From Fig. 8a,
we have the following relation between electron abundance and $R_1$ in between 100AU to 6000 AU.
$$
\frac{n_e}{n_H}= 3.2502 \times 10^{-6} -4.0302 \times 10^{-6} \ R_1.
$$
In the dense interior (100AU - 6000AU), $R_2$ should be used to measure the degree of ionization. 
From our model, this is in the range of $10^{-9} -10^{-6}$ in this region. It has
the following relation (Fig. 8b):
$$
\frac{n_e}{n_H} = 6.3773 \times 10^{-6} -0.00012136 \ R_2 +0.00015693 \ R_2^2
$$
In reality, ionization degree depends on various parameters. R$_1$ and R$_2$ could be used to
diagonize it to some extent. Since, abundance of any species depends on various
physical conditions and age of the source, it is very difficult to present any generalized 
formula for predicting the ionization degree at any particular instant.

\begin{figure}
\vskip 2cm
\centering{
\vbox{
\includegraphics[height=8cm,width=12cm]{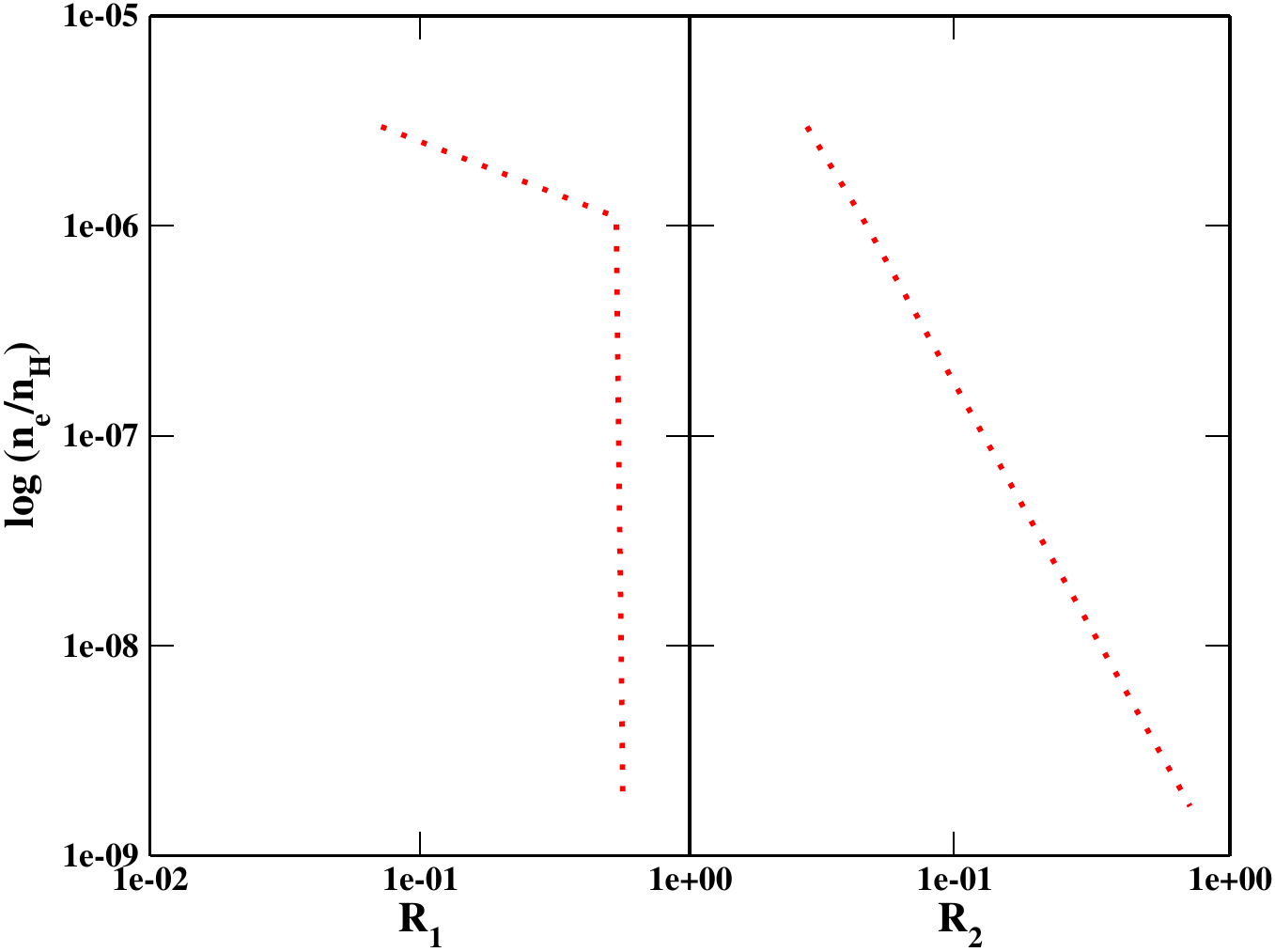}}}
\caption{\small Electron abundance relative to (a) $R_1$ and (b) $R_2$.}
\end{figure}

So far, a number of theoretical models tried to correlate physical properties
of protostars with time since gravitational collapse begins (e.g., Smith, 1998; Myers et al., 1998). 
Time sequences of evolutionary stages determined by various models are more or less similar
but absolute ages vary significantly (Emprechitinger et al., 2008 and references therein). 
According to Froebrich (2005), absolute age for Class 0/I borderline objects, for example, 
varies between 10$^4$ years and a few times $10^5$ years.
According to Emprechtinger et al. (2008), R$_2$ is known to trace the
evolution of prestellar cores. They proposed that R$_2$ could be used to trace core
evolution even after the star formation. According to them, protostars with R$_2>0.5$
are in a stage shortly after beginning of collapse. Later on, R$_2$ decreases 
until it reaches a value of $0.03$ at Class 0/I borderline. 

Several deuterated molecules are found to be heavily abundant in ISM.
Even water could also be highly fractionated. HDO could efficiently be formed on the grain
surfaces by the deuteration reaction. Stark et al. (2004) reported detection of
HDO ground transition towards IRAS 16293 and derived a HDO abundance of $\sim$ 10$^{-10}$
in cold region of the envelope.
Measurement of water deuterium fractionation is a relevant tool for understanding mechanisms of
water formation and evolution from prestellar phase to formation of planets and comets
(Coutens et al., 2013). Several attempts were made to
derive HDO/H$_2$O ratio for various Class 0 protostars, which correspond to main
accretion phase. But results turned out to be quite different from one another. Measurements of HDO/H$_2$O
were carried out at two separate zones. For T$>100$ K, region of
IRAS 16293-2422, Parise et al. (2005) and Coutens et al. (2012, 2013) estimated a HDO/H$_2$O ratio about
a few percent, using single dish observations, whereas Persson et al. (2013) found much lower
estimate ($~ 9 \times 10^{-4}$) using interferometric data. For $T<100$ K region, the ratio of HDO/H$_2$O 
found to be between $3 \times 10^{-3} -1.5 \times 10^{-2}$ in IRAS 16293-2422 by Coutens et al. (2012, 2013).

A large amount of doubly deuterated formaldehyde (D$_2$CO) has been observed in solar type
proto-star IRAS 16293-2422 (Ceccarelli et al., 1998). 
Turner (1990) found that D$_2$CO/H$_2$CO $\sim 0.003$ in the
Orion compact ridge. It was expected that this high
fractionation of D$_2$CO occurs on the grain during the cold phase and 
that the species are evaporated to gas component during warm phase. 
Moreover, a detection of doubly deuterated and triply deuterated methanol 
was reported by Parise et al. (2002, 2004) towards low mass protostar IRAS 16293-2422.

\begin{figure}
\vskip 2cm
\centering{
\vbox{
\includegraphics[height=12cm,width=8cm]{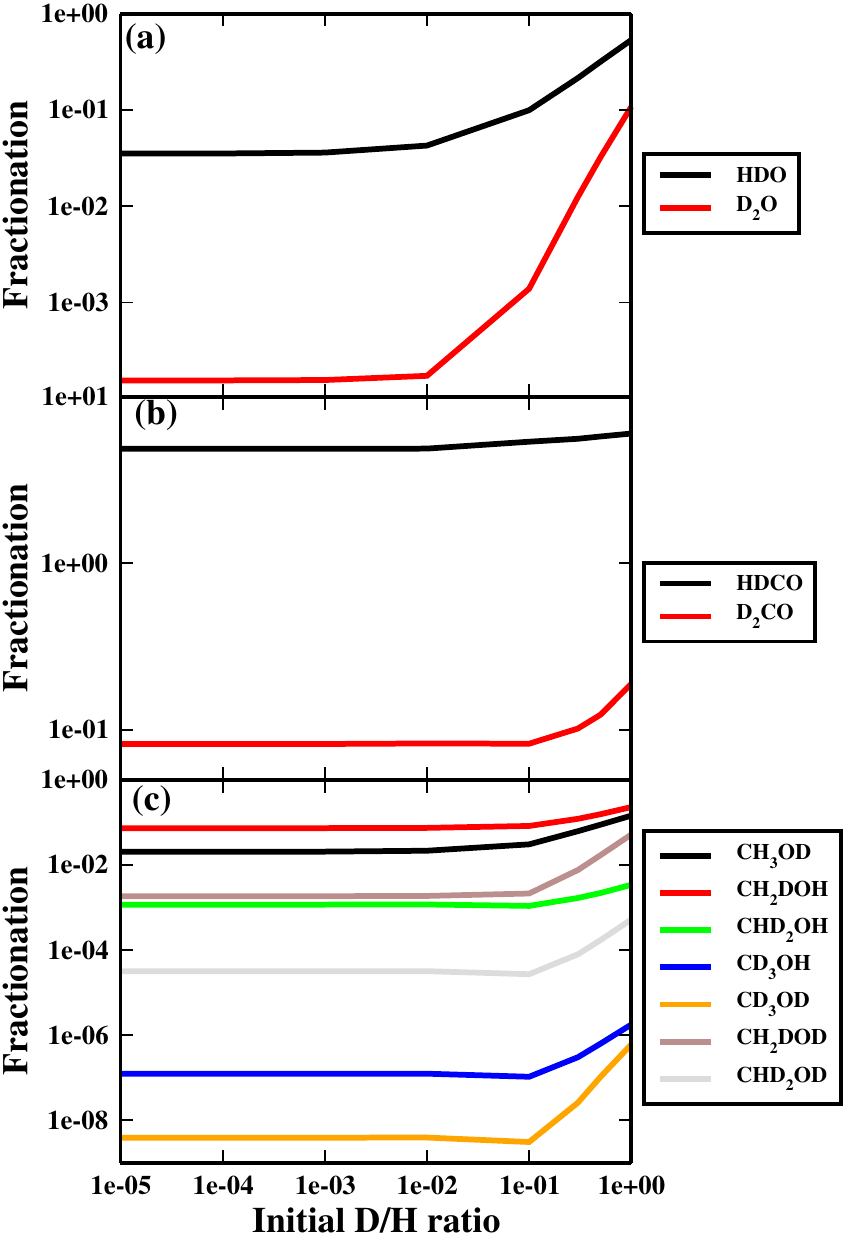}}}
\caption{\small Deuterium fractionation of Ice phase (a) H$_2$O, (b) H$_2$CO and (c) CH$_3$OH molecules.}
\end{figure}

In Fig. 9abc, we show deuterium fractionation ratio of some of the major ice phase species with variation of initial atomic D/H ratio of gas phase. 
In the study from Roberts et al. (2003) the D/H ratio can reach $0.3$.
Here, we vary the initial atomic D/H ratio from $10^{-5}$ to $1$.
For this case, we consider set 1 energy values, 
$n_H= 10^4$ cm$^{-3}$, T=10K, A$_V =10$ and vary initial 
atomic D/H ratio. Deuterium fractionation values are taken during the late time 
(beyond $\sim 10^6$ year).
Figure 9abc clearly shows that singly deuterated part of H$_2$O, H$_2$CO and CH$_3$OH
are heavily fractionated. HDCO fractionation crossing unity for most of the time. 
We notice that during the late time (beyond $10^6$ year), abundance of HDCO is enhanced over 
the abundance of $H_2$CO. In Caselli et al. (2002b), it was assumed that activation barriers for the
reactions such as H+CO ($E_1$) and H+H$_2$CO ($E_2$) were in the range of $1000-2000$ K. For the same reaction
with deuterium, i.e., D+CO activation barrier differ slightly due to 
zero point vibrations. They assumed for the reaction D+CO, the barrier is $E_1-70 \ K$.
For reactions D+H$_2$CO, H+HDCO, D+HDCO, H+D$_2$CO and D+D$_2$CO energy barriers were
considered to be $E_2-201$ K, $E_2-35$ K, $E_2-242$ K, $E_2-75$ K and $E_2-287$ K respectively. In this
paper, unless otherwise stated, we used this consideration alone and assumed 
$E_1=E_2=2000$ K. Most recent findings by Fuchs et al. (2009), suggest that the energy 
barriers ($E_1$ \& E$_2$) could be much lower than what was suggested by Caselli et al. (2002b).
According to Fuchs et al. (2009), for the reaction H+CO, activation energy barrier
is $390 \pm 40$ K and for the reaction H+H$_2$CO it is $415 \pm 40$ K.
In a test case, we have considered $E_1=390$ K and $E_2=415$ K respectively for H+CO and H+H$_2$CO
reactions. Keeping in mind the consideration of Caselli et al. (2002b), activation 
energy barriers for D+CO, D+H$_2$CO, H+HDCO, D+HDCO, H+D$_2$CO and D+D$_2$CO reactions 
become $320$ K, $214$ K, $380$ K, $173$ K, $340$ K and $128$ K respectively.
In reality, the difference in zero point vibrations should depend on the size of the barrier 
and for lower barriers these differences should be less. Here, we are considering the 
barrier energies which were considered by Caselli et al. (2002) and references therein. 
They assumed that for the reactions in which D atoms partially replace the H atoms have 
different activation energy barriers than $E_1$ and $E_2$ due to zero-point vibrations. 
These values were calculated  by D. Woon (private communication), these differences are accurate, 
regardless of the accuracy of the barriers for the ${\mathrm H + CO}$ and ${\mathrm H + H_2CO}$ 
reactions. According to Kaiser et al. (1999), ab-initio calculations show that isotopic 
substitution of H by D influences zero- point vibrational energy. Thus,
differences in activation energy barriers for above reactions are obvious. 
Isotopic substitution can modify the rate of reactions in various ways. In many cases, 
differences can be rationalized by noting how the mass of an atom affects 
vibrational frequency of a chemical bond even if the electron configuration is 
nearly identical. Heavier atoms will (classically) lead to lower vibration frequencies, 
or, viewed quantum mechanically, will have lower zero-point energy. With a lower 
zero-point energy, more energy must be supplied to break the bond, 
resulting in a higher activation energy for bond cleavage, which in turn lowers the measured rate
(Carten et al., 2011; Carten et al., 2012). 
 
We ran our code for all these sets (set 1, set 2, set 3 and experimental activation barrier using set 1). 
To have an idea of how the results differ, in Table 3,
we present fractionation ratios along with their respective column densities 
for initial atomic D/H ratio $0.1$, n$_H=10^4$, $T=10$ K and A$_V=10$.
Moreover, in Table 3, we include the same for various energy values. 
Since for ice phase species, we reach a steady state during late stage of the evolution
process, column densities and fractionation ratios are given in 
Table 3 during late stages (after the simulation time is over $\sim 10^7$ year) of simulation.

\section{Quantum Chemical Model}
\subsection{Computational details}
A significant amount of work is reported in the literature where spectroscopic 
investigation of different interstellar species is guided by theoretical predictions 
(e.g., Huang \& Lee, 2008, 2009; Majumdar et al., 2013, 2014ab; Das et al., 2013a).  
Vibrational spectroscopy and rotational spectroscopy are widely used to identify
several species of interstellar interest. To compute
vibrational (harmonic) frequencies, we optimize geometries of different deuterated neutrals and
ions at density functional theory based on Becke three parameter Exchange and Lee, Yang and Parr 
correlation functional (B3LYP, Becke 1993, Lee, Yang \& Parr, 1988) with 6-311++G basis set. 
We use B3LYP/6-311++G levels of theory for optimization of 
different interstellar species. Geometry optimization enabled us to locate minimum 
energy configuration of these deuterated 
species. This procedure calculates wave function and energy at a starting geometry and 
then proceeds to search for a new geometry of a lower energy. This is repeated until the lowest 
energy geometry of these species is found. The procedure calculates forces on each atom
of species by evaluating gradient of energy with respect to atomic positions. 
Vibrational frequencies for all these species are also obtained at the 
same level of theory and these frequencies depend on second derivative of energy 
with respect to nuclear positions. To study chemical as well as spectroscopic 
properties of all neutral deuterated species in ice phase (water ice), 
we first optimize geometry of these species at B3LYP/6-311++G level with 
integral equation formalism variant 
(IEFPCM) as default self-consistent reaction field (SCRF) method with a dielectric constant of $78.5$. 
SCRF calculation using IEFPCM model as implemented in Gaussian 09W (Frisch et al., 2009) program was used 
to include bulk solvation effect of medium as water ice. 
The IEFPCM bulk solvent medium is simulated as a continuum of dielectric constant ($=78.5$), 
which surrounds a solute cavity defined by the union of a series of interlocking spheres 
centered on the atoms. For this computation, we optimize geometries of different 
neutral deuterated species at B3LYP/6-311++G level of theory of Gaussian 09W program with the 
polarizable continuum model and the integral equation formalism variant as a default SCRF method. 
These methods provide a proper ice phase environment for these neutral deuterated condensed phase species. 
Here, solute-solvent electrostatic interactions are treated at dipole level. 
Solvent effect brings significant changes in geometrical parameters of these condensed 
phase deuterated species. Our model confirms that polarization of the solute by continuum 
has important effects on absolute and relative solvation energies, which, in turn shift
frequency as compared to gas phase. 

To the best of our knowledge, no detailed investigation 
has been carried out for computation of different rotational and centrifugal-distortional 
constants for different deuterated interstellar neutral and ions simultaneously. 
Usually, these constants are evaluated at the same level as corresponding force fields needed 
either for prediction of vibrational frequencies or vibrational corrections to 
rotational constants. Here, we use MP2/6-311++G(d,p) level of theory for performing our 
calculations. Corrections for interaction between rotational motion and vibrational motion 
along with corrections for vibrational averaging and anharmonic corrections to vibrational 
motion are also considered in our calculations. 
In brief, we report here rotational and distortional constants of all important 
neutral and ionic 
deuterated species in our model, which are corrected for each vibrational state as well as 
vibrationally averaged structures. These rotational constants are required to predict 
different rotational transitions and this can be done using the `SPCAT' program (Pickett, 1991). 

\subsection{Results of Quantum chemical modeling}
Several gas phase deuterated molecules are observed till date mostly by observing
rotational transitions. We carried out quantum chemical calculations
by using the method described in section 3.1. 

A complex pattern of chemical change could be visible due to collapse of a dense cloud core leading to the
formation of a young star and its circumstellar disk. Processes such as depletion
of molecules on cold icy grains during collapsing phase, evaporation of newly-formed
species when a protostar starts to heat its surroundings, and high temperature reactions
in shocked zones created by impacts of outflow, cycle molecules from one compound
into another. These changes are not only of chemical interest, but can also be used as diagnostics of
physical state of evolution of an object. Moreover, knowledge of chemistry is needed to
choose proper molecular line to trace physical structure of a particular component. In
single-dish observations with $15"-30"$ beams, all of these different chemical processes are
blurred together, whereas current interferometers with a few arcsec resolution suffer from poor spatial
sampling. ALMA, with its unprecedented sensitivity, resolution and UV coverage will be necessary to zoom in
and image these different chemical regimes and quantitatively address chemical evolution in the initial
stages of star formation. Due to this reason, we concentrate only on $3$ to $9$ ($84-720$GHz) band of ALMA.
In order to assist observational identifications, in Table 4, we provide different 
rotational parameters for the most important neutral gas phase deuterated species. 
These rotational parameters are necessary to predict spectrum of these species 
(Pickett, 1991). Among different neutral deuterated species in Table 4, DNC and DCN 
are linear molecules. In order to get all rotational parameters for these species, 
we need to apply degenerate perturbation theory to correctly compute bending modes of 
these species. But as this feature is not implemented in Gaussian 09W program, 
we obtain only single rotational constant (B) for DCN and DNC.
Moreover, some rotational transitions which 
are falling in  $84-720$GHz (bands $3-9$ of ALMA) range, are also given in Table 5. 
Column density of deuterated species are calculated
by using the following relation used by Shalabiea et al. (1994), Das \& Chakrabarti (2011) and Das et al. (2013a).
$$
N(A) = n_H x_i R,
$$
where, $n_H$ is the total hydrogen number density, $x_i$ is the abundance of $i^{th}$ species and $R$ 
is path length along the line of sight ($= 1.6 \times 10^{21} A_V )/n_H$).
Peak column density of all the neutral deuterated species are presented in Table 5. 
Observed column densities are also given with corresponding references.
Peak values of ice phase column densities are tabulated along with gas phase
column densities. In case of formaldehyde and methanol, our calculated gas phase column densities 
differ by some orders of magnitudes 
from the observed values. Our calculated ice phase column densities might be a clue for
this. For example, as per Loinard et al. (2000), observed column density of HDCO varies
between $4.8 \times 10^{13} -8.1 \times 10^{13}$ cm$^{-2}$. Our calculated peak value of the
gas phase HDCO is $3.23 \times 10^8$ cm$^{-2}$. But peak value of calculated ice phase
column density is $6.48 \times 10^{15}$ which indicates that these types of species are mainly
synthesized on dust surfaces at low temperature and could populate gas phase subsequently via some
energetic events (stellar ejecta, in hot cores associated with proto-stars, in dense photo-dissociation 
regions associated with luminous stars, or in the post-shock regions). 
So, HDCO could be produced during the cold phase of the molecular cloud but when the 
surrounding cloud become warmer (by some energetic events), ice would sublime and release all the species in the gas phase. We compare our 
theoretically computed transitions with earlier works and 
find that some of the transitions match exactly. Calculated column density of 
deuterated isotopomers of methanol and H$_2$CO are found to be very low. These molecules
could efficiently be formed on grain surfaces and could populate the gas phase by some energetic events.

As in the case of gas phase neutral deuterated species, here also 
different rotational and distortional constants for deuterated ions 
are given in Table 6. To obtain all rotational and distortional constants 
for linear molecules, such as  DCO$^+$, DCN$^+$, CD$^+$ and N$_2$D$^+$, it is necessary to use
degenerate perturbation theory which could correctly account for bending modes.  
But this feature is not implemented in GAUSSIAN 09W program. That is why we tabulate only 
`B' value for these species.
Peak column densities of these gas phase deuterated ions along with its possible 
rotational transitions only in the range of $84-720$GHz are presented in Table 6. A comparison with our
computed transitions are also highlighted in Table 7.

In order to describe inconsistencies between transitions reported in this work and other
catalogs (JPL, NIST), we would like to mention that first step towards observation of new molecules
in an ISM is to obtain transition frequencies.  
For this, rotational and distortional constants have to be calculated using a suitable method.
Hence it is always important and advantageous to have a good frequency prediction before an experiment or
observation as spectrometers do not generally have automated scanning mode. This 
significantly reduce manual scanning time in high resolution spectroscopic observations. 
Rotational and distortional constants can be calculated using two methods, namely, by computational
quantum chemistry and experiments. In order to measure experimental rotational spectra of different
species at temperatures around $10$ K or less (interstellar environment), one can use cavity Fourier transform,
chirped pulse or with free jets, stark modulation spectrometers etc. Millimeter wave ($30-300$ GHz) and
sub-millimeter wave ($300-3000$ GHz) rotational spectra are most efficiently obtained in frequency
domain with broad-band spectrometers. Expected resolution of the rotational spectrum is
determined by Doppler limited line widths of absorption lines but sometimes 
also by experimental techniques used for the detection. Now, most of rotational and distortional
constants in a catalogs are reported by fitting (using the SPFIT program
package written by H. M. Pickett of JPL) observed experimental transitions
(using the above methods) with selected Hamiltonian. SPFIT program provides extended
possibilities for definition of molecular Hamiltonian required for a particular problem.
This program also allows declaration of spectroscopic parameters for all type of rotors. The
fitting program produces an input file with improved spectroscopic constants for
use of SPCAT program to calculate the spectra.
Computational chemistry has greatly enhanced predictive power for such 
experiments and observations. For most of the molecular system, exact solution of the
Schr\"{o}inger equation are not possible. Molecular system that are studied here ranges
from neutral deuterated to ionic deuterated species. The smallest
neutral system studied in this work is HDO with 3 atoms and 11 electrons. {\it Ab initio} method
MP2 was used in this work which scale as $N^5$ in computational time with the number of electrons.
This method is proven to be accurate as compared to other 
available methods (Das et al., 2013a; Majumdar et al., 2014a; Majumdar et al., 2014b).
The goal of calculations performed prior to an experiment is to generate coordinates for
the nuclei in a molecule. Moment of inertia can be obtained from coordinates
of atoms and masses. First, center of mass is determined and coordinates
are then generated from origins placed at the center of mass. The principle axes of the
system can be determined by diagonalizing moment of inertia matrix. Once moment of
inertia is determined for the principle axis, rotational constants can be
determined. Symmetric and asymmetric top spectra are known combinations of rotational
constants. The SPCAT program then can be used to generate spectrum and transitions from
rotational and distortional constants. This prediction will be the starting point for
all the molecules reported in this work.

Majumdar et al. (2014a) reported rotational and distortional constants for CH$_2$CN$^-$, CHDCN$^-$
and CD$_2$CN$^-$ in symmetrically reduced Hamiltonian.
Experimental values of these constants were available for CH$_2$CN$^-$ and C$D_2$CN$^-$
by fitting observed experimental transitions with Watson S-reduced Hamiltonian using a least
square routine (Lykke et al., 1987). Majumdar et al. (2014a) mentioned that errors on computed
line frequencies are related to errors on calculated rotational and distortional constants.
There were some uncertainties from experimentally obtained values as well. From there, Majumdar et al. (2014a)
pointed out that these uncertainties could result in an error in between $0.6$MHz and $12$MHz for
frequency range in between $18$GHz and $319$GHz. Higher uncertainty was associated with 
higher frequencies. Here, we observe a similar situation. Tables 5 and 7 clearly show that our
results are in agreement at lower frequencies ($<200$GHz). 
Our predicted frequencies for deuterated formaldehyde and methanol are within the 
error bar of $2$ GHz for transitions below $300$ GHz. In case of D$_2$CO, 
our calculated transition for $4_{04}\rightarrow 3_{03}$ is at $231.988$ GHz whereas
in JPL catalog this transition is tabulated at $231.41021$ GHz, so the error bar is $0.578$ GHz.
In case of CH$_2$DOH our calculated transition for $2_{02}\rightarrow 1_{01}$
is at $89.80$ GHz whereas in JPL catalog this transition is pointed out at $89.25$ GHz
and for CH$_3$OD our calculated transition for $2_{02}\rightarrow 1_{01}$
is at $91.3$ GHz whereas in JPL catalog this transition is pointed out at $90.703$ GHz
In JPL or NIST, no frequencies for our given transitions are presented
for CH$_2$DOD, CHD$_2$OD, CD$_3$OD. Since our computed transitions for
singly and doubly deuterated methanols are in good agreement for the lower frequency range ($<300GHz$), 
we hope our computed frequncies will be helpful to detect CH$_2$DOD, CHD$_2$OD, CD$_3$OD in the ISM.
These frequencies along with our computed column densities could be used to predict the 
source antenna temperature by using CASSIS interactive spectrum analyzer 
(Caux et al., 2011).  Now to convert the antenna temperature to the source flux density, 
following relation could be used:
$$
S_{\nu} =3520 \times \frac{T_a}{\eta_A D^2},
$$
where, $\eta_A$  is the the aperture efficiency and $D$ is the diameter. By including all the values, 
one could have the source flux density ($S_{\nu}$). 
This information would be extremely helpful for the observer for the detection of these species in the ISM.

Theoretical prediction of rotational and distortional constants 
depends on various choices (Bowman et al., 2007; Carter et al., 2009):
i. Choice of ab initio potentials (depend on degrees of freedom),
ii. Choice of ab initio dipole moment surface,
iii. Development of essential quantum methods to calculate rovibrational wave functions.
First of all, it is very time consuming and each species requires different quantum 
chemical treatments. Computational studies could be feasible for smaller molecular species, 
extensions to larger molecular systems with various variable (e.g. mass, charge for ions etc.) 
remain a computational challenge. Currently, full spectral treatment for molecules having more 
than few atoms is limited to the use of effective Hamiltonians, which often do not provide 
results close to observations or experiments (Carter et al., 2011, 2012). 
Theoretical computed spectral line-lists provide an essential complement to experimentally measured 
ones. One obvious way that theory can complement experiment is in filling in the gaps in 
laboratory-measurements. Theoretical line-lists can in principle span the  entire spectral 
range. 

In earlier Section, it was pointed out that various deuterated molecules could be trapped 
inside an interstellar ice. Existence of deuterated isotopomers in interstellar 
ice analogues could be traced by observing through IR telescopes. In order to assist 
observers, we calculate vibrational frequencies for various deuterated isotopomers. 
For this computation, we optimize geometries of molecules by density functional theory based on 
B3LYP method and 6-311++G basis set of GAUSSIAN 09W program. In Table 8, IR features of
various ice phase isotopomers of Water, H$_2$CO and CH$_3$OH are given. We have also tabulated
gas phase IR features of these isotopomers to highlight spectral changes due to isotopic substitution.
In Table 9, we provide IR features of gas phase
deuterated ions. Only for representation, in Fig. 10a and Fig. 10b, we show
IR spectra of gas phase DCO$^+$, HCO$^+$, N$_2$D$^+$ and N$_2$H$^+$ respectively. 
This Figure explains that isotopic substitution plays a dominant role in IR spectral features.
\begin{figure}
\vskip 2cm
\centering{
\vbox{
\includegraphics[height=7cm,width=8cm]{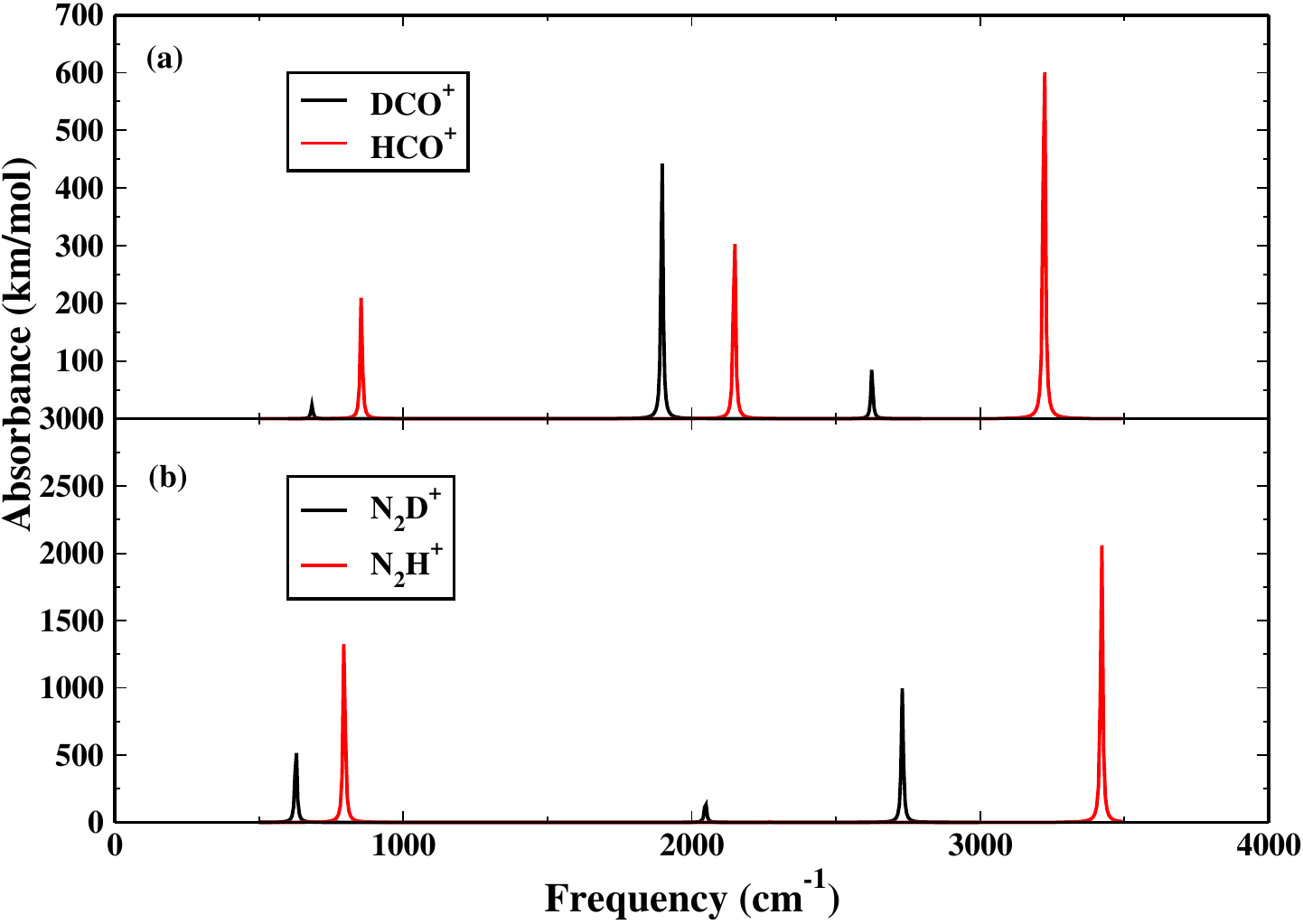}}}
\caption{\small IR spectra of gas phase (a) HCO$^+$ and DCO$^+$ and (b) N$_2$H$^+$ and N$_2$D$^+$.}
\end{figure}

\section{Conclusions}
Despite a low elemental abundance of atomic deuterium in interstellar space,
some species are observed to be heavily fractionated by deuterium. Sometimes fractionation ratio crosses 
observed elemental D/H ratio ($\sim 1.5 \times 10^{-5}$). In order to find out
evolution history of deuterated species, we make use of a large 
gas-grain chemical network. We applied our model for mimicking cold dark
cloud region of an ISM. Moreover, we have also performed quantum chemical simulation to compute 
several spectral parameters which could encourage observers for further investigation
of deuterated species in and around an ISM. 
In brief, followings are the highlights of the present work:

$\bullet$Chemical evolutions of deuterated molecules in gas and ice phases are discussed.
Computed abundances of gas phase deuterated isotopomers of formaldehyde and some isotopomers of
methanol are found to be below present observational limits. However, their abundances in ice phase
are reasonable. We hope that these ice phase molecules could populate the gas phase by some
energetic events making them detectable.

$\bullet$Column densities of deuterated species are computed and are compared with
observed column densities.

$\bullet$Rotational and distortional constants for some of the important deuterated 
species are computed by quantum chemical simulations.

$\bullet$Rotational and vibrational (harmonic) spectral transitions are calculated for some of the 
abundant deuterated species and are compared with other theoretical, observational or
experimental databases.

$\bullet$Dependences of deuterium fractionation of ice phase species for different 
binding energies and activation energy barriers are discussed.

$\bullet$ Various regions of a collapsing cloud are modeled
and radial distribution of R$_1$ (DCO$^+$/HCO$^+$) and R$_2$ (N$_2$D$^+$/N$_2$H$^+$) is studied.

\subsection{Acknowledgments}
AD and DS are grateful to ISRO for financial support through a respond project
(Grant No. ISRO/RES/2/372/11-12), SKC acknowledges a DST project
(Grant No.SR/S2/HEP-40/2008) for partial financial support and LM thanks
MOES project for his partial financial support. Authors would like to thank an 
anonymous referee whose valuable suggestions helped to improve this paper significantly.

{}

\clearpage
\newpage

}
\end{table}

\clearpage

\begin{table}
\addtolength{\tabcolsep}{-4pt}
\caption{Calculated and observed rotational transitions of some neutral gas phase deuterated species}
{\scriptsize
\centering
\begin{tabular}{|c|c|c|c|c|c|}
\hline
{\bf Neutral deuterated species}&{\bf Gas phase/Ice phase}&{\bf Observed/estimated}&{\bf Transitions}&{\bf Calculated frequency} & {\bf Comparison with}\\
&{\bf Column density }&{\bf column density}&{} & {\bf (in GHz)}& {\bf (Observation/Theory)} \\
&{\bf (in cm$^{-2}$)}&{(\bf in cm$^{-2})$}&&&\\
\hline
\hline
{\bf HDO} &  $1.19 \times 10^{15}/1.62 \times 10^{17}$ &$4.4 \times 10^{13}$$^{p9}$ &  1$_{01}$-0$_{00}$ & 470.854 &464.924$^{ni}$   \\
& &&  2$_{02}$-1$_{01}$ & 941.432 & 919.3109$^{n,j}$\\
\hline
{\bf D$_2$O} &  $1.13 \times 10^{14}/2.26 \times 10^{15}$ &$2.5 \times 10^{12}$$^c$&  1$_{01}$-0$_{00}$ & 368.807    &    \\
&& &  2$_{02}$-1$_{01}$ & 737.390 & \\
\hline
&&   &   2$_{02}$-1$_{01}$ &  141.753 & 144.8280015$^{p9}$ \\
\hline
\end{tabular}}
\end{table}

\clearpage

{\scriptsize
\centering
\addtolength{\tabcolsep}{-4pt}
\begin{tabular}{|c|c|c|c|c|c|}
\hline
{\bf Neutral deuterated species}&{\bf Gas phase}&{\bf Observed/estimated}&{\bf Transitions}&{\bf Calculated frequency} & {\bf Comparison with}\\
&{\bf Column density }&{\bf column density}&{} & {\bf (in GHz)}& {\bf (Observation/Theory)} \\
&{\bf (in cm$^{-2}$)}&{(\bf in cm$^{-2})$}&&&\\
\hline
\hline
& &   &   2$_{02}$-1$_{01}$ &  84.512 & \\
& &  &   3$_{03}$-2$_{02}$ &  126.766 & \\
& &   &   4$_{04}$-3$_{03}$ &  169.018 & \\
& &   &   5$_{05}$-4$_{04}$ &  211.266 & \\
{\bf CH$_2$DOD} &  $5.71 \times 10^{12}/7.73\times 10^{14}$ && 6$_{06}$-5$_{05}$ &  253.510 & \\
\hline
\end{tabular}

\clearpage

\begin{table}
\addtolength{\tabcolsep}{-2pt}
\caption{Rotational and distortional constants for different deuterated ions at MP2/6-311G++(d,p) level of theory}
{\scriptsize
\centering
\begin{tabular}{|c|c|c|c|c|}
\hline
{\bf Gas phase deuterated species}&{\bf Rotational constants}&{\bf Calculated values (in MHz)}&
{\bf Distortional constants}&{\bf Calculated values (in MHz)} \\
\hline
\hline
&A&  -    & $D_J$&   - \\
{\bf DCO$^+$}& B &  35483.057 &$D_{JK}$& - \\
&C& -  &$D_{K}$& -    \\
& & & $d_1$& - \\
&&& $d_2$& - \\
\hline
&A&  -  &   $D_J$&   - \\
{\bf N$_2$D$^+$}& B &  37491.716 &$D_{JK}$& - \\
\hline
\end{tabular}}
\end{table}

\clearpage

\begin{table}
\caption{Calculated and observed rotational transitions of some deuterated ions}
{\scriptsize
\centering
\begin{tabular}{|c|c|c|c|c|c|}
\hline
{\bf Important deuterated ions}&{\bf Gas Phase }&{\bf Observed/estimated}&{\bf Transitions}&{\bf Calculated frequency} & {\bf Comparison}\\
{}&{\bf Column density }&{\bf Column density}&{} & {\bf (in GHz)}& {\bf (Obs./Theo.)} \\
&{\bf  (in cm$^{-2}$) }&{\bf (in cm$^{-2}$)}&&&\\
\hline
\hline
&& & 1$_{01}$-0$_{00}$ & 70.966 & 72.039$^b,j,ni$\\
&&   &   2$_{02}$-1$_{01}$ &  141.932 &144.077$^{b,ni}$ \\
& &  &   3$_{03}$-2$_{02}$ &  212.898 & 216.112$^{b,ni}$\\
& &   &   4$_{04}$-3$_{03}$ &  283.864 & 288.143$^{o,ni}$ \\
& &   &   5$_{05}$-4$_{04}$ &  354.830 & 360.169$^{ni}$\\
\hline
\end{tabular}}
\end{table}

\clearpage

\begin{table}
\caption{\bf Vibrational frequencies of different isotopomers in gas/ice phase at B3LYP/6-311++G level of theory}
{\scriptsize
\centering
\begin{tabular}{|c|c|c|c|c|c|c|}
\hline
{\bf Deuterated species}&{\bf Isotopomers}&{\bf Peak} & {\bf Absorbance} & {\bf Peak }&{\bf Absorbance}&{\bf Experimental}\\
{\bf in ice Phase}&{ }&{\bf positions }&{}&{\bf positions }&{} & {\bf peak} \\
&&\bf ( in cm$^{-1}$)& &\bf ( in cm$^{-1}$)& & \bf positions (in gas phase) \\
{}&{}& {\bf (Gas phase)}&{}&{\bf (H$_2$O ice)}&&{\bf( in cm$^{-1}$)}\\
\hline
\hline
&&  1378.16& 88.6531  &  1405.66 &  122.7328 & \\
{\bf H$_2$O}& {\bf HDO}& 2705.63& 7.2636   &  2702.10 & 15.7450 & \\
&& 3735.97& 19.1116&   3727.40 & 41.5054 & \\
\cline{2-7}
&& 1154.12  & 54.8490    &  1176.99 &  76.6104 & \\
\end{tabular}}
\end{table}

\clearpage

{\scriptsize
\centering
\begin{tabular}{|c|c|c|c|c|c|c|}
\hline
{\bf Deuterated species}&{\bf Isotopomers}&{\bf Peak} & {\bf Absorbance} & {\bf Peak }&{\bf Absorbance}&{\bf Experimental}\\
{\bf in ice Phase}&{ }&{\bf positions }&{}&{\bf positions }&{} & {\bf peak} \\
&&\bf ( in cm$^{-1}$)& &\bf ( in cm$^{-1}$)& & \bf positions \\
{}&{}& {\bf (Gas phase)}&{}&{\bf (H$_2$O ice)}&&{\bf( in cm$^{-1}$)}\\
\hline
&&  169.30    & 92.7351  &      296.56 & 259.9902 & \\
{\bf CH$_3$OH}&& 384.91  & 106.3536  &       905.41 & 2.0358 & \\
&& 404.02  &  146.5793  &   951.71 & 149.3625 & \\
&& 596.94  & 93.4742  &      1058.02 &9.8248 & \\
\hline
\end{tabular}}

\clearpage

\begin{table}
\caption{Vibrational frequencies of different deuterated ions in gas phase at B3LYP/6-311++G level of theory}
{\scriptsize
\centering
\vbox{
\begin{tabular}{|c|c|c|c|}
\hline
{\bf Gas phase deuterated ions}&{\bf Peak }&{\bf Absorbance}\\
{deuterated ions}&{\bf positions }&{} \\
{}& {\bf (gas phase)}&{}\\
&\bf (Wavenumber)&   \\
&\bf ( in cm$^{-1}$)&  \\
\hline
\hline
 &  683.57   &  3.3998    \\
{\bf DCO$^+$}&   1897.22 &  132.8371  \\
& 2624.76 & 25.3446  \\
\hline
&  628.33   &  87.6580    \\
{\bf N$_2$D$^+$}&   2047.85 &  49.4619  \\
& 2730.37 & 290.2813  \\
\hline
&  2230.79  &  66.6525    \\
{\bf H$_2$D$^+$}&   2345.31 &  317.0829  \\
& 3015.87 & 76.1241  \\
\hline
&  1946.29  &  158.8752   \\
{\bf HD$_2$$^+$}&   2098.01 &  87.1753  \\
& 2733.51 & 126.7451  \\
\hline
&  566.82  &  856.4178   \\
&   1415.30 &  118.6293  \\
{\bf H$_2$DO$^+$}& 1695.32 & 188.8344  \\
& 2675.50 & 145.2111  \\
& 3639.63& 201.6169  \\
& 3742.37 & 504.7184  \\
\hline
&  518.63  &  685.2860   \\
&   1246.29 &  79.5791  \\
{\bf HD$_2$O$^+$}& 1557.51 & 149.7999  \\
& 2599.43 &  59.0382   \\
& 2760.07 & 265.0375  \\
& 3694.16 & 364.9741  \\
\hline
&  313.84  &  3.8689   \\
&   838.27 & 94.7313  \\
{\bf HDCN$^+$}& 970.70 & 131.1481  \\
& 1781.07 &  12.5791   \\
& 2139.99 & 211.2602  \\
& 2867.97 & 378.5943  \\
\hline
&  288.00  &  1.1228   \\
&   742.61 & 51.1535  \\
{\bf D$_2$CN$^+$}& 802.03 & 76.3573  \\
& 1737.90 & 0.3969   \\
& 2116.60 & 159.4864  \\
& 2157.11 & 265.8839  \\
\hline
&  465.38  &  515.2134   \\
&   1241.35 & 77.8866  \\
{\bf D$_3$O$^+$}& 1241.36 & 77.8855  \\
& 2529.44 & 0.0000   \\
& 2759.90 & 260.9641  \\
& 2759.90 & 260.9620  \\
\hline
{\bf CD$^+$} & 2011.51 & 0.9199  \\
\hline
&  248.79  &  4.8847   \\
&   705.87 & 2.3408  \\
{\bf C$_3$D$^+$}& 711.54 & 3.0668  \\
& 1158.20 & 54.3460   \\
& 2022.23 & 834.8943  \\
& 2605.23 & 46.2760  \\
\hline
&   617.26 & 28.3187  \\
{\bf DCN$^+$}& 1943.59 & 6.7620 \\
& 2714.08 & 26.1128   \\
\hline
\end{tabular}}}
\end{table}
\clearpage

{\Large{Corrigendum}}
\vskip 0.5cm

$\rm{Ankan Das^{1},\ Liton Majumdar^{2},\ Sandip K. Chakrabarti^{1}, Dipen Sahu^{3}}$\\
\vskip 0.5cm

$^1$Indian Centre For Space Physics, 43 Chalantika, Garia Station Road, Kolkata 700084, India,\\

$^2$Jet Propulsion Laboratory, California Institute of Technology, 4800 Oak Grove Drive, Pasadena, CA 91109, USA,\\

$^3$Indian Institute of Astrophysics, Sarjapur Main Road, 2nd Block, Koramangala, Bangalore-560034, India.\\ 

\vskip 1cm

{\justifying \large {Recently we have found that some of the spectroscopic data published in our manuscript `Deuterium Enrichment of the Interstellar Medium' 
were wrongly calculated. These data were presented in Table 4, 5, 6, 7, 8, and 9 of the original manuscript. We would like to add a corrigendum
to correct these mistakes. Also we would like like to acknowledge Milan Sil and Prasanta Gorai for providing the corrected spectroscopic data.
The inclusion of the new data does not change any conclusion of the paper but should be helpful for the future use in the scientific community. 
Again, we would like to apologize for the mistake and any inconvenience cause.}}}

\vskip 5cm 

-----------------------------------------------------------------\\
DOI of original article: $<$10.1016/j.newast.2014.07.006$>$\\
$<$Corresponding author details$>$ Ankan Das\\
$<$Corresponding author email address$>$ ankan.das@gmail.com
\clearpage

\begin{table}
\scriptsize{
\centering
{  Updated Table 4: Rotational and distortional constants for different neutral deuterated species at MP2/6-311G++(d,p) level of theory}\\
\vskip 0.5 cm
\vbox{
\begin{tabular}{|c|c|c|c|c|c|}
\hline
{  Sl. No.} &{  Gas phase}&{  Rotational}&{  Calculated values (in MHz)}&
{  Distortional}&{  Calculated values (in MHz)} \\
 & {  deuterated species} & {  constants} & & {  constants} &  \\
\hline
\hline
&&A&   {  678954.0862529} &   $D_N$ & {  0.11489973$\times$10$^{2}$} \\
& & B &  {  275116.2887393} &    $D_{K}$ & {  0.56862951$\times$10$^{3}$} \\
1. & {  HDO (Singlet)}& C & {  195783.5955611} &  $D_{NK}$ & {  -0.42717512$\times$10$^{2}$}  \\
& && & $d_N$ & {  0.39942106$\times$10$^{1}$} \\
&&&& $d_K$ & {  0.14005582$\times$10$^{2}$} \\
\hline
&&A&   {  445069.0163982}   & $D_N$ & {  0.93617494$\times$10$^{1}$} \\
\hline
\end{tabular}}}
\end{table}

\begin{table}
\vskip -3.0 cm
{\scriptsize
\addtolength{\tabcolsep}{-4pt}
{  Updated Table 5: Calculated and observed rotational transitions of some neutral gas phase deuterated species}
\vskip 0.5 cm
\centering
\begin{tabular}{|c|c|c|c|c|c|}
\hline
{  Neutral deuterated species}&{  Gas phase/Ice phase}&{  Observed/estimated}&{  Transitions}&{  Calculated frequency} & {  Comparison with}\\
&{  Column density }&{  column density}&{} & {  (in GHz)}& {  (Observation/Theory)} \\
&{  (in cm$^{-2}$)}&{(  in cm$^{-2})$}&&&\\
\hline
\hline
{  HDO} &  $1.19 \times 10^{15}/1.62 \times 10^{17}$ &$4.4 \times 10^{13}$$^{p9}$ &  1$_{01}$-0$_{00}$ & 470.854 &464.924$^{ni}$   \\
& &&  2$_{02}$-1$_{01}$ & 941.432 & 919.3109$^{n,j}$\\
\hline
{  D$_2$O} &  $1.13 \times 10^{14}/2.26 \times 10^{15}$ &$2.5 \times 10^{12}$$^c$&  1$_{01}$-0$_{00}$ & 368.807    &    \\
&& &  2$_{02}$-1$_{01}$ & 737.390 & \\
\hline
&&   &   2$_{02}$-1$_{01}$ &  141.753 & 144.8280015$^{p9}$ \\
\hline
\end{tabular}}
\end{table}

\clearpage

{\scriptsize
\centering
\addtolength{\tabcolsep}{-4pt}
\begin{tabular}{|c|c|c|c|c|c|}
\hline
{  Neutral deuterated species}&{  Gas phase}&{  Observed/estimated}&{  Transitions}&{  Calculated frequency} & {  Comparison with}\\
&{  Column density }&{  column density}&{} & {  (in GHz)}& {  (Observation/Theory)} \\
&{  (in cm$^{-2}$)}&{(  in cm$^{-2})$}&&&\\
\hline
\hline
& &   &   2$_{02}$-1$_{01}$ &  {  84.476} & \\
& &  &   3$_{03}$-2$_{02}$ &  {  126.713} & \\
& &   &   4$_{04}$-3$_{03}$ &  {  168.949} & \\
& &   &   5$_{05}$-4$_{04}$ &  {  211.183} & \\
{  CH$_2$DOD} &  $5.71 \times 10^{12}/7.73\times 10^{14}$ && 6$_{06}$-5$_{05}$ &  {  253.414} & \\
\hline
\end{tabular}


\begin{table}
\scriptsize{
\centering
{  Updated Table 6: Rotational and distortional constants for different deuterated ions at MP2/6-311G++(d,p) level of theory}
\vskip 0.5 cm
\vbox{
\begin{tabular}{|c|c|c|c|c|c|}
\hline
{  Sl. No.} &{  Gas phase}&{  Rotational}&{  Calculated values (in MHz)}&
{  Distortional}&{  Calculated values (in MHz)} \\
& {  deuterated species} & {  constants} &  & {  constants} & \\
\hline
\hline
&&A&  -    & $D_N$&   - \\
&& B &  {  35483.0567432} &$D_{K}$& - \\
1. & {  DCO$^+$ (Singlet)}&C& {  35483.0567432} &$D_{NK}$& -    \\
& & && $d_N$& - \\
&&&& $d_K$& - \\
\hline
&&A&  -  &   $D_N$&   - \\
\hline
\end{tabular}}}
\end{table}

\clearpage

\begin{table}
{\scriptsize
\vskip -3.3 cm
{  Updated Table 7: Calculated and observed rotational transitions of some deuterated ions}
\vskip 0.5 cm
\centering
\begin{tabular}{|c|c|c|c|c|c|}
\hline
{  Gas phase deuterated ions}&{  Gas Phase }&{  Observed/estimated}&{  Transitions}&{  Calculated frequency} & {  Comparison}\\
{}&{  Column density }&{  Column density}&{} & {  (in GHz)}& {  (Obs./Theo.)} \\
&{   (in cm$^{-2}$) }&{  (in cm$^{-2}$)}&&&\\
\hline
\hline
&& & 1$_{01}$-0$_{00}$ & 70.966 & 72.039$^b,j,ni$\\
&&   &   2$_{02}$-1$_{01}$ &  141.932 &144.077$^{b,ni}$ \\
& &  &   3$_{03}$-2$_{02}$ &  212.898 & 216.112$^{b,ni}$\\
& &   &   4$_{04}$-3$_{03}$ &  283.864 & 288.143$^{o,ni}$ \\
& &   &   5$_{05}$-4$_{04}$ &  354.830 & 360.169$^{ni}$\\
\hline
\end{tabular}}
\end{table}

\begin{table}
\scriptsize{
\vskip -2.4 cm
{  Updated Table 8: Vibrational frequencies of different isotopomers in gas/ice phase at B3LYP/6-311++G level of theory}
\vskip 0.5cm
\centering
\begin{tabular}{|c|c|c|c|c|c|c|}
\hline
{  Species}&{  Deuterated isotopomers}&{  Peak} & {  Absorbance} & {  Peak }&{  Absorbance}&{  Experimental}\\
& &{  positions }&{}&{  positions }&{} & {  peak} \\
&&  ( in cm$^{-1}$)& &  ( in cm$^{-1}$)& &   positions (in gas phase) \\
{}&{}& {  (Gas phase)}&{}&{  (H$_2$O ice)}&&{ ( in cm$^{-1}$)}\\
\hline
\hline
&&  1378.16& 88.6531  &  1405.66 &  122.7328 & \\
{  H$_2$O}& {  HDO}& 2705.63& 7.2636   &  2702.10 & 15.7450 & \\
&& 3735.97& 19.1116&   3727.40 & 41.5054 & \\
\cline{2-7}
&& 1154.12  & 54.8490    &  1176.99 &  76.6104 & \\
\end{tabular}}
\end{table}

\clearpage

{\scriptsize
\centering
\begin{tabular}{|c|c|c|c|c|c|c|}
\hline
{  Species}&{  Deuterated isotopomers}&{  Peak} & {  Absorbance} & {  Peak }&{  Absorbance}&{  Experimental}\\
&&{  positions }&{}&{  positions }&{} & {  peak} \\
&&  (in cm$^{-1}$)& &  (in cm$^{-1}$)& &   positions \\
{}&{}& {  (Gas phase)}&{}&{  (H$_2$O ice)}&&{ ( in cm$^{-1}$)}\\
\hline
& & {  286.90} & {  201.1437} & {  284.23} & {  269.0283} & \\
& & {  845.10} & {  33.8673} & {  845.15} & {  47.4969} & \\
& & {  882.72} & {  1.0519} & {  884.02} & {  3.0216} & \\
& & {  945.44} & {  50.7212} & {  928.46} & {  72.6139} & \\
\hline
\end{tabular}}

\begin{table}
\scriptsize{
{  Updated Table 9: Vibrational frequencies of different deuterated ions in gas phase at B3LYP/6-311++G level of theory}
\vskip 0.5 cm
\centering
\vbox{
\begin{tabular}{|c|c|c|c|c|}
\hline
{  Sl. No.} & {  Gas phase deuterated ions}&{  Peak positions }&{  Absorbance}\\
& &{   (Wavenumber)}&   \\
& & {  (in cm$^{-1}$)}& \\
\hline
\hline
& &  683.57   &  3.3998    \\
& & {  683.57} & {  3.3998} \\
1. & {  DCO$^+$ (Singlet)}&   1897.22 &  132.8371  \\
& & 2624.76 & 25.3446  \\
\hline
& &  628.33   &  87.6580    \\
\hline
\end{tabular}}}
\end{table}

\end{document}